\documentclass{aa}

\usepackage[varg]{txfonts}
\usepackage{epsfig,graphicx,natbib,url,twoopt}
\usepackage[varg]{txfonts}
\usepackage{hyperref}          
\hypersetup{
 colorlinks=true,  
 urlcolor=blue,    
 linkcolor=red,     
 citecolor=blue 
}

\setcitestyle{citesep={,}} 

\def\kms{\hbox{km$\;$s$^{-1}$}}
      
\def\mAA{m\AA}
\def\Halpha{\mbox{H\hspace{0.1ex}$\alpha$}}
\def\Hbeta{\mbox{H\hspace{0.1ex}$\beta$}}

\def\FeI{\ion{Fe}{i}}
\def\FeIline{\ion{Fe}{i}~6173\,\AA}

\def\CaK{\ion{Ca}{ii}~K}

\defcitealias{2020A&A...641L...5J}{Paper I}

\begin{document}

\title{Properties of ubiquitous magnetic reconnection events in the lower solar atmosphere}

\author {Jayant Joshi\inst{1,2,3} 
\and
Luc H. M. Rouppe van der Voort\inst{2,3} 
}
\authorrunning{Joshi \& Rouppe van der Voort}
\titlerunning{Properties of reconnection events in the lower solar atmosphere}

\institute{Indian Institute of Astrophysics, 
    II Block, Koramangala, Bengaluru 560 034, India
\and
  Institute of Theoretical Astrophysics,
  University of Oslo, %
  P.O. Box 1029 Blindern, N-0315 Oslo, Norway
\and
  Rosseland Centre for Solar Physics,
  University of Oslo, %
  P.O. Box 1029 Blindern, N-0315 Oslo, Norway\\
  \email{jayant.joshi@iiap.res.in}
  }


\abstract
{
Magnetic reconnection in the deep solar atmosphere can give rise to enhanced emission in the Balmer hydrogen lines, a phenomenon referred to as Ellerman bombs. Recent high quality \Hbeta{} observations indicate that Ellerman bombs are more common than previously thought and it was estimated that at any time about half a million Ellerman bombs are present in the quiet Sun. 
}
{
We performed an extensive statistical characterization of the quiet Sun Ellerman bombs (QSEBs) in these new \Hbeta{} observations. 
}
{
We analyzed a 1~h dataset of quiet Sun observed with the Swedish 1-m Solar Telescope that consists of spectral imaging in the \Hbeta\ and \Halpha\ lines, as well as spectropolarimetric imaging in \FeIline{}.
We used the $k$-means clustering and the 3D connected component labeling techniques to automatically detect QSEBs.
}
{
We detected a total of 2809 QSEBs. 
The lifetime varies between 9~s and 20.5~min with a median of 1.14~min. 
The maximum area ranges between 0.0016 and 0.2603~Mm$^2$ with a median of 0.018~Mm$^2$.
The maximum brightness in the \Hbeta\ wing varies between 1.06 and 2.76 with respect to the average wing intensity.
A subset (14\%) of the QSEBs display enhancement of the \Hbeta\ line core. On average, the line core brightening appears 0.88~min after the onset of brightening in the wings, and the distance between these brightenings is 243~km. This gives rise to an apparent propagation speed ranging between $-$14.3 and +23.5~\kms, with an average that is upward propagating at +4.4~\kms. The average orientation is nearly parallel to the limbward direction.
QSEBs are nearly uniformly distributed over the field of view but we find empty areas with the size of mesogranulation.
QSEBs are located more frequent near the magnetic network where they are often bigger, longer lived and brighter. 
} 
{
We conclude that QSEBs are ubiquitous in quiet Sun and appear everywhere except in areas of mesogranular size with weakest magnetic field ($B_{\rm{LOS}}\lesssim50$~G). Our observations support the interpretation of reconnection along vertically extended current sheets. 
}

\maketitle

\begin{figure*}[!ht]
\centering
\includegraphics[width=\textwidth]{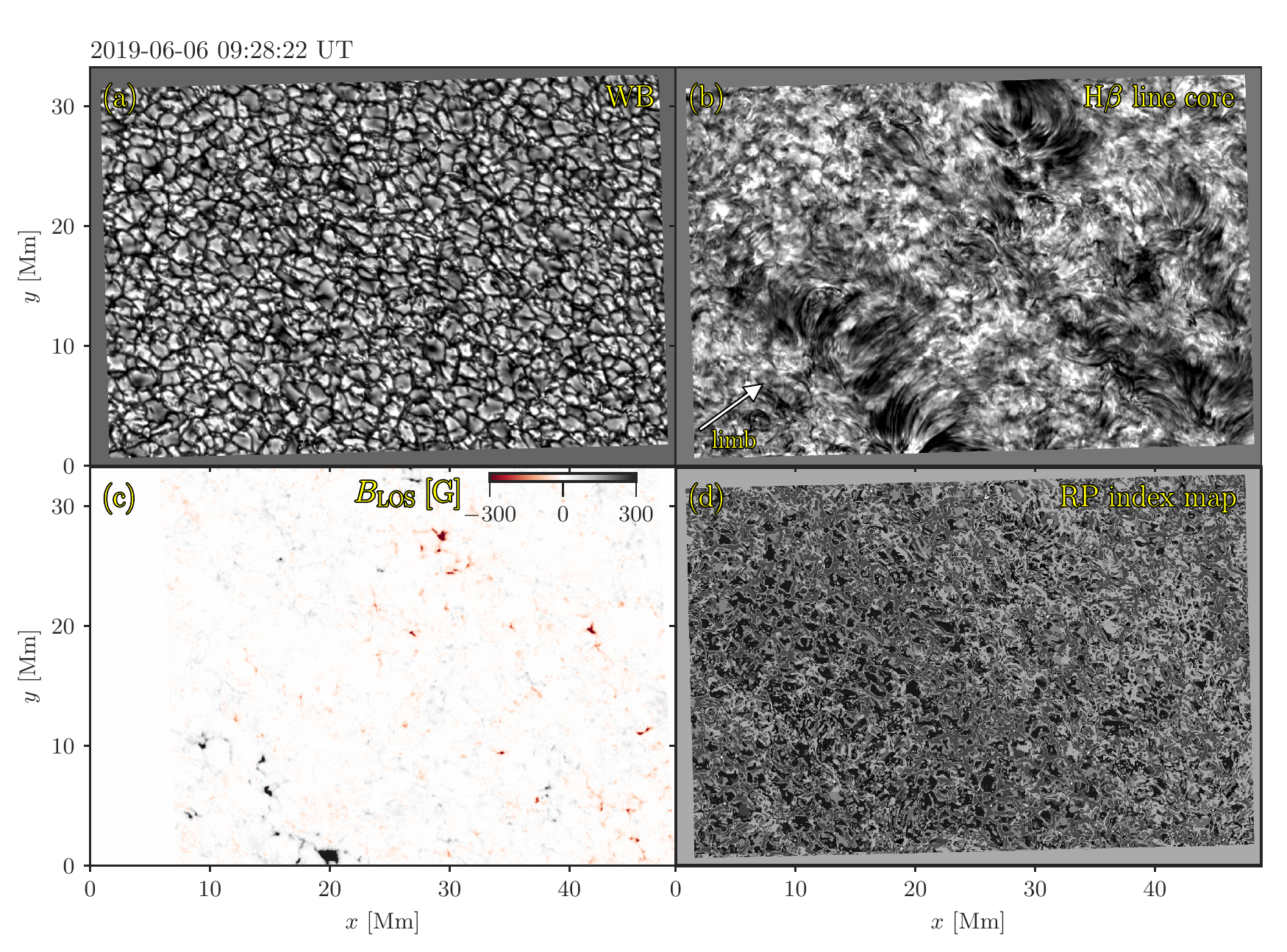}
\caption{\label{fig:overview}%
Quiet Sun FOV observed by SST on 6 June 2019. (a) CHROMIS WB image, (b) \Hbeta{} line core image, (c) line-of-sight (LOS) magnetic field (saturated at $\pm300$~G) retrieved from ME inversions of the \FeI{}~6173~\AA\ line observed with CRISP. Note that the FOV for CRISP is smaller than CHROMIS so that there is no overlap for 8~Mm at left. (d) RP index map obtained from the \textit{k}-means clustering algorithm applied to the \Hbeta{} line profiles showing all 100 RPs in shades of grey. The arrow in panel (b) shows the direction towards the nearest limb.
An animation of this figure is available \href{https://www.mn.uio.no/astro/english/people/aca/rouppe/movies/joshi_qseb2022_fig01.mp4}{online}.
}
\end{figure*}

\section{Introduction}
\label{sec:introduction}

The effect of magnetic reconnection on the solar atmosphere can be observed over a wide range of spatial and temporal scales. 
This ranges from flares and eruptions on the scale of active regions down to small, short-lived brightenings in the intergranular lanes in the deep solar atmosphere. 
These small-scale events are referred to as Ellerman bombs 
\citep[EBs,][]{1917ApJ....46..298E} 
when observed as enhancements of the broad spectral wings of the hydrogen Balmer lines. 
These EBs are most clearly observed in active regions with strong flux emergence where they are located around the polarity inversion line and appear as subarcsecond sized brightenings in the \Halpha\ or \Hbeta\ line wing 
\citep[see, e.g.,][]{2002ApJ...575..506G, 
2004ApJ...614.1099P, 
2006ApJ...643.1325F, 
2007A&A...473..279P, 
2008PASJ...60..577M, 
2008ApJ...684..736W, 
2017A&A...598A..33L}. 
%
The characteristic Balmer EB spectral profile is traditionally referred to as moustache like
\citep{1964ARA&A...2..363S}: 
enhanced wings that appear in emission (with peak emission around a Doppler offset of 40~\kms) and line core absorption that has similar low intensity level as the surroundings. 
%
High-resolution imaging spectroscopy has indicated that the enhanced wing emission can be attributed to heating in the low atmosphere and that the reconnection site is effectively obscured by the overlying chromospheric canopy of fibrils in the \Halpha\ line core
\citep{2011ApJ...736...71W, 
2013ApJ...774...32V, 
2013ApJ...779..125N}. 
The interpretation of EBs being a sub-canopy phenomenon is supported by recent ALMA observations 
\citep{2020A&A...643A..41D} 
and numerical modelling
\citep{2017ApJ...839...22H, 
2017A&A...601A.122D, 
2019A&A...626A..33H}. 

Under inclined observing angle, \Halpha\ wing images show EBs as tiny (1--2~Mm), bright, upright flames that flicker rapidly on a time scale of seconds 
\citep{2011ApJ...736...71W, 
2013JPhCS.440a2007R, 
2015ApJ...798...19N}. 
A wide spread in EB lifetimes is reported in the literature, for example, 
\citet{1973SoPh...30..449R} 
and
\citet{1982SoPh...79...77K} 
reported average lifetimes between 11 and 13~min, with the longest living more than 40 min. 
In a sample of 139 EBs detected in high spatial resolution \Halpha\ observations, 
\citet{2013ApJ...774...32V} 
found that 75\%\ had lifetimes less than 5~min. 

The traditional view that the EB phenomenon is exclusive for active regions was challenged when first
\citet{2016A&A...592A.100R} 
and later 
\citet{2017ApJ...845...16N} 
and \citet{2018MNRAS.479.3274S} 
observed tiny ($\lesssim0\farcs5$) Ellerman-like brightenings in quiet Sun when observed at extremely high spatial resolution.
Recently, \citet[][hereafter \citetalias{2020A&A...641L...5J}]{2020A&A...641L...5J} 
analysed new \Hbeta\ observations and found that quiet Sun EBs (QSEB) are much more ubiquitous than the earlier \Halpha\ observations suggested. 
The shorter wavelength of the \Hbeta\ line allowed for higher spatial resolution and larger contrast and facilitated detection of smaller and weaker EB events. 
The analysis suggested that about half a million QSEBs are present in the solar atmosphere at any time. 
The ubiquity of QSEBs raises the question of the contribution of small-scale magnetic reconnection events on the total energy budget of the solar atmosphere. 

In this paper, we present extensive analysis of the observations used in 
\citetalias{2020A&A...641L...5J}. 
Whereas the analysis in 
\citetalias{2020A&A...641L...5J} 
was primarily concentrated on the best seeing samples, we here analyze the full time sequence. 
We present a detailed discussion of the detection method and statistics on QSEB properties like area, lifetime and brightness. 
We find a strong correlation between number of QSEB detections and seeing quality.
The QSEB phenomenon could be one of the prime motivations to strive for higher spatial resolution in solar physics. 

\section{Observations}
\label{sec:obs}

\begin{figure*}[!ht]
\includegraphics[width=\textwidth]{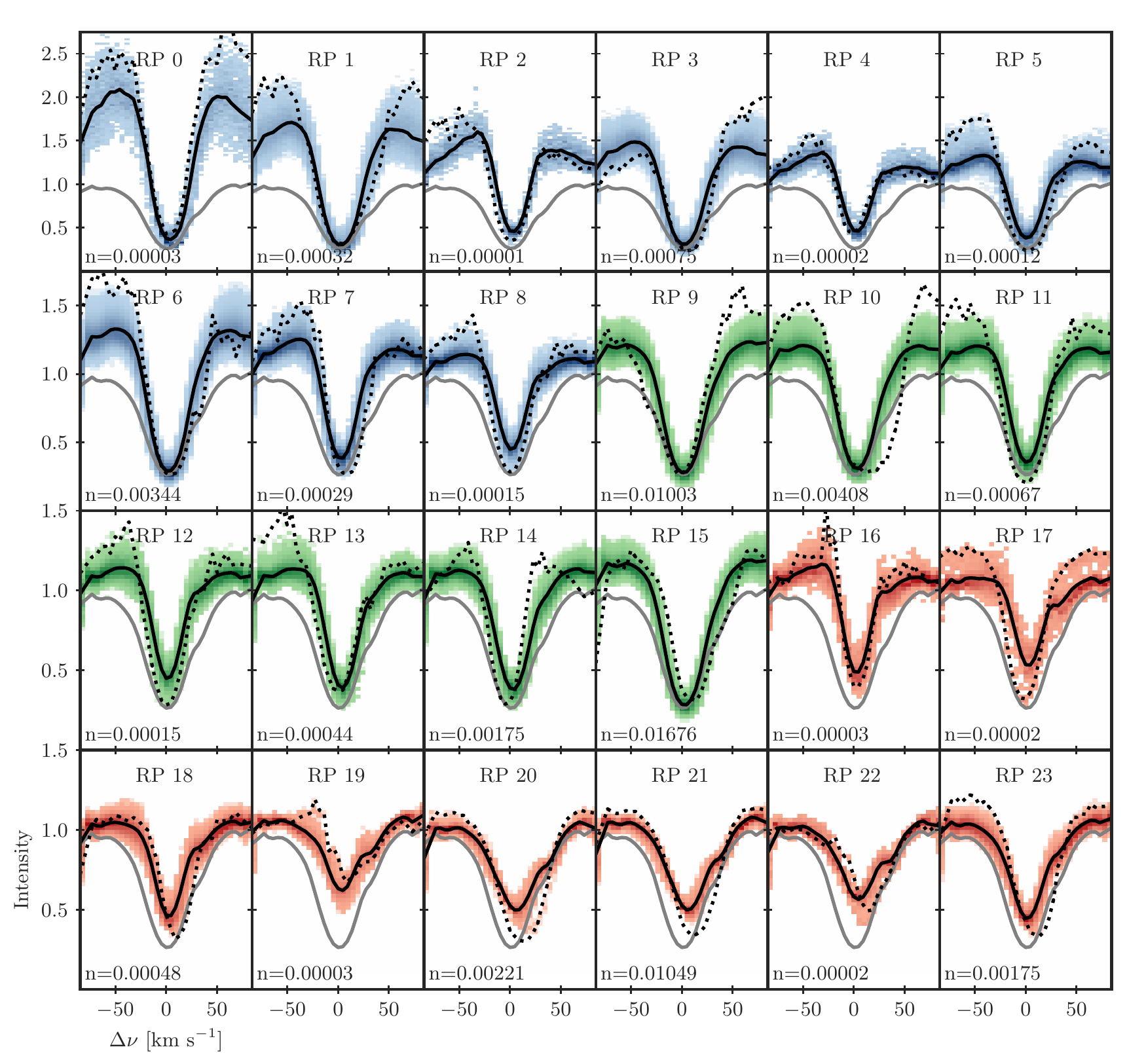}
\caption{\label{fig:RPs}%
Twenty four representative profiles (RPs) from the $k$-means clustering of the \Hbeta{} line that are identified as signature of QSEB. The black lines show RPs whereas shaded colored areas represent density distribution of \Hbeta{} spectra within a cluster; darker shades indicate higher density. Within a particular cluster, the \Hbeta{} profile that is farthest (measured in euclidean distance) from the corresponding RPs is shown by the black dotted line. As reference, the average quiet Sun profile (gray line) is plotted in each panel. RPs 0--8 show the typical EB-like \Hbeta{} profiles, i.e., significantly enhanced wings and unaffected line core, while RPs 9--15 display weak enhancement in the wings. RPs 15--23 show intensity enhancement in the line core. The parameter $n$ represents the number of pixels in a cluster as percentage 
of the total of $\sim3.07\times10^{10}$ pixels.    
}
\end{figure*}

\begin{figure*}[!ht]
\centering
\includegraphics[width=\textwidth]{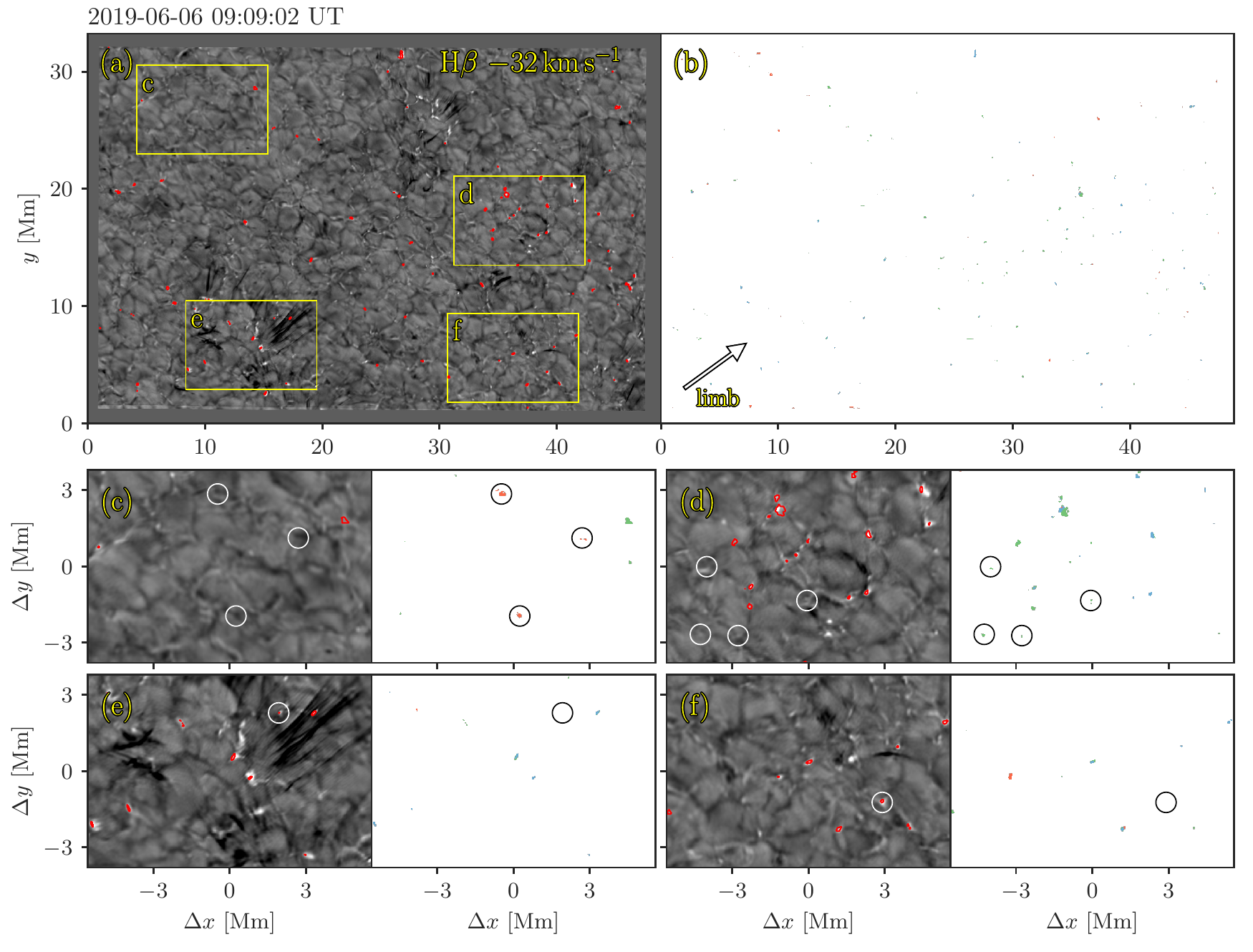}
\caption{\label{fig:kmeans_label}%
Detections of QSEBs using the \textit{k}-means clustering technique and morphological operations.
Panel (a) shows the observed FOV in the \Hbeta{} blue wing at Doppler offset $-32$~km~s$^{-1}$. The red contours mark locations of 91 detected QSEBs. Panel (b) shows the locations of the selected QSEB RPs shown in Fig.~\ref{fig:RPs}; locations of RP 0--8, RP 9--15, and RP 16--23 are indicated by blue, green, and red colours, respectively. Panels (c--f) show zoom-ins on four different areas marked by the white boxes in (a) showing a similar pair of panels of \Hbeta\ wing and RP maps as in (a) and (b).
Circles in panels (c) and (d) mark examples of areas of RPs that did not end up as QSEB detections: these areas could not be connected in space or time to nearby RP 0--8 locations. Circles in (e) and (f) mark examples of QSEB detections that do not have selected RPs in this particular map: these detections were connected to RP 0--8 locations in the time steps before or after.  The arrow in panel (b) shows the direction towards the nearest limb.
An animation of this figure is available \href{https://www.mn.uio.no/astro/english/people/aca/rouppe/movies/joshi_qseb2022_fig03.mp4}{online}
}
\end{figure*}

\begin{figure*}[!ht]
\centering
\includegraphics[width=\textwidth]{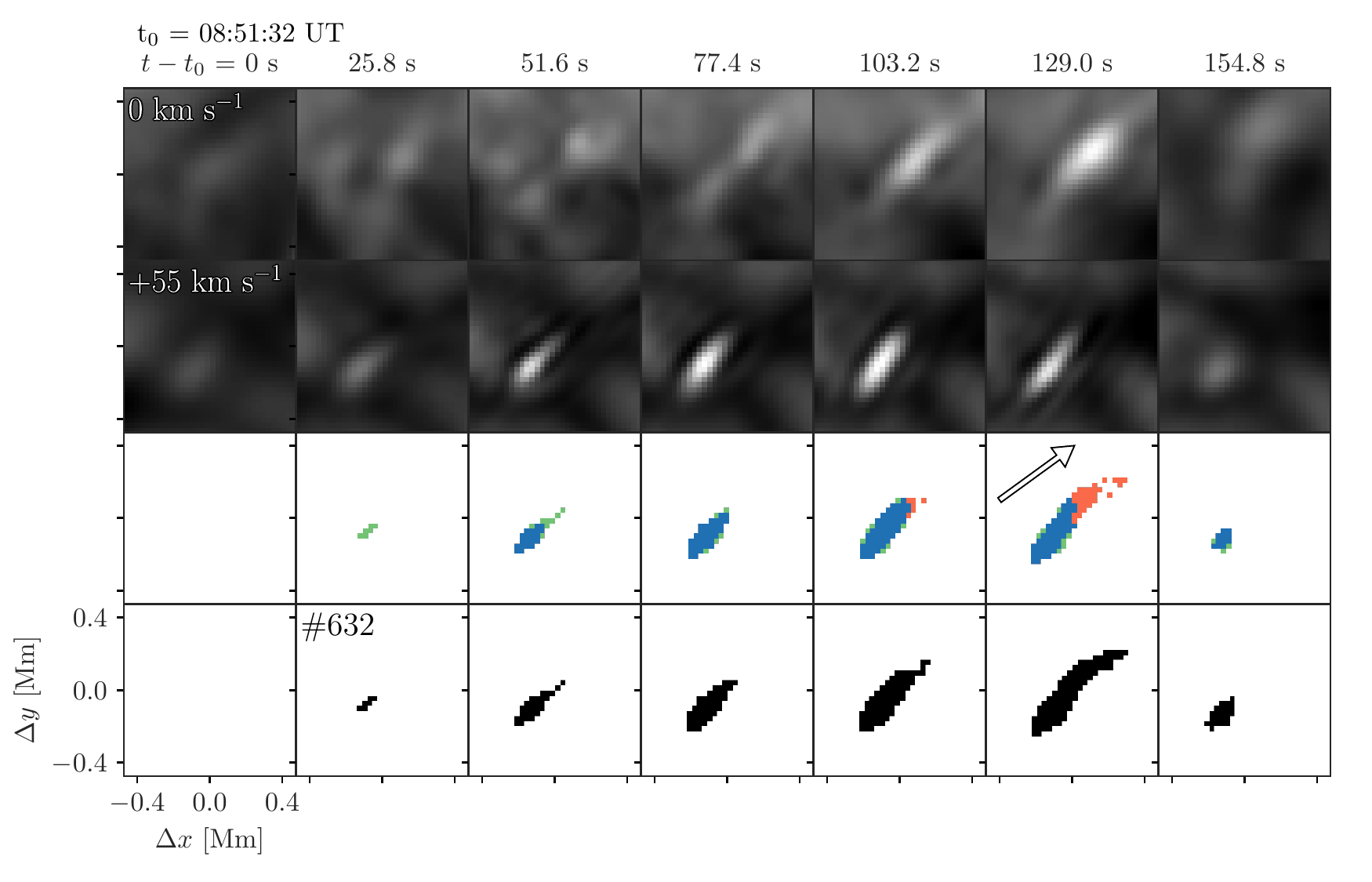}
\caption{\label{fig:evolution}%
Temporal evolution of a QSEB and illustration of the detection method. The top row shows a series of \Hbeta{} line core images, every third image is shown as the cadence is 8.6~s. The second row shows the corresponding \Hbeta{} red wing images. The third row shows the locations of the selected QSEB RPs shown in Fig.~\ref{fig:RPs}; locations of RP 0--8, RP 9--15, and RP 16--23 are indicated by blue, green, and red colours, respectively. The bottom rows shows the corresponding binary masks of QSEB detections after the morphological operation. The arrow in the third row shows the direction towards the nearest limb.
}
\end{figure*}


\begin{figure}[!ht]
\centering
\includegraphics[width=0.48\textwidth]{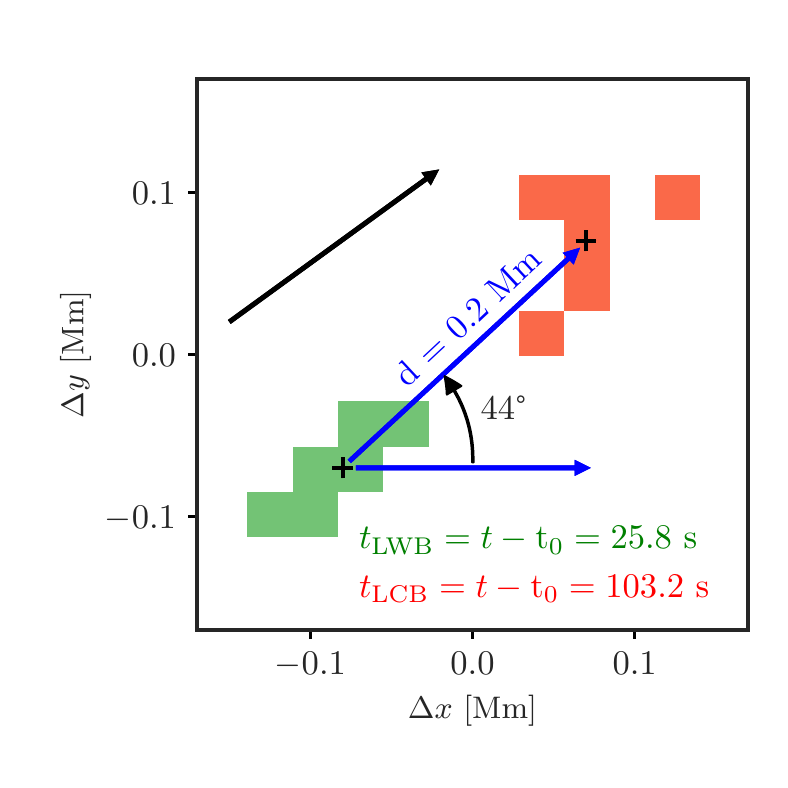}
\caption{\label{fig:vel_est}%
Measurement of line core brightening (LCB) with respect to line wing brightening (LWB) in the \Hbeta{} line for the QSEB shown in Fig.~\ref{fig:evolution}. Distance $d$ between LWB and LCB is the separation between the centers of gravity of the areas with RPs 9--15 (LWB, green) and RPs 16--23 (LCB, red) at the time of their first appearances. 
The time separation between these first appearances (77.4~s) is used to determine the average propagation speed, here 2.6~\kms.
The orientation is measured against the horizontal direction as shown, the direction towards the nearest limb is 36$\degr$ (black arrow). The histograms of measurements of all QSEBs with line core brightening are shown in Fig.~\ref{fig:lcb_stats}. 
}
\end{figure}


\begin{figure*}[!ht]
\includegraphics[width=\textwidth]{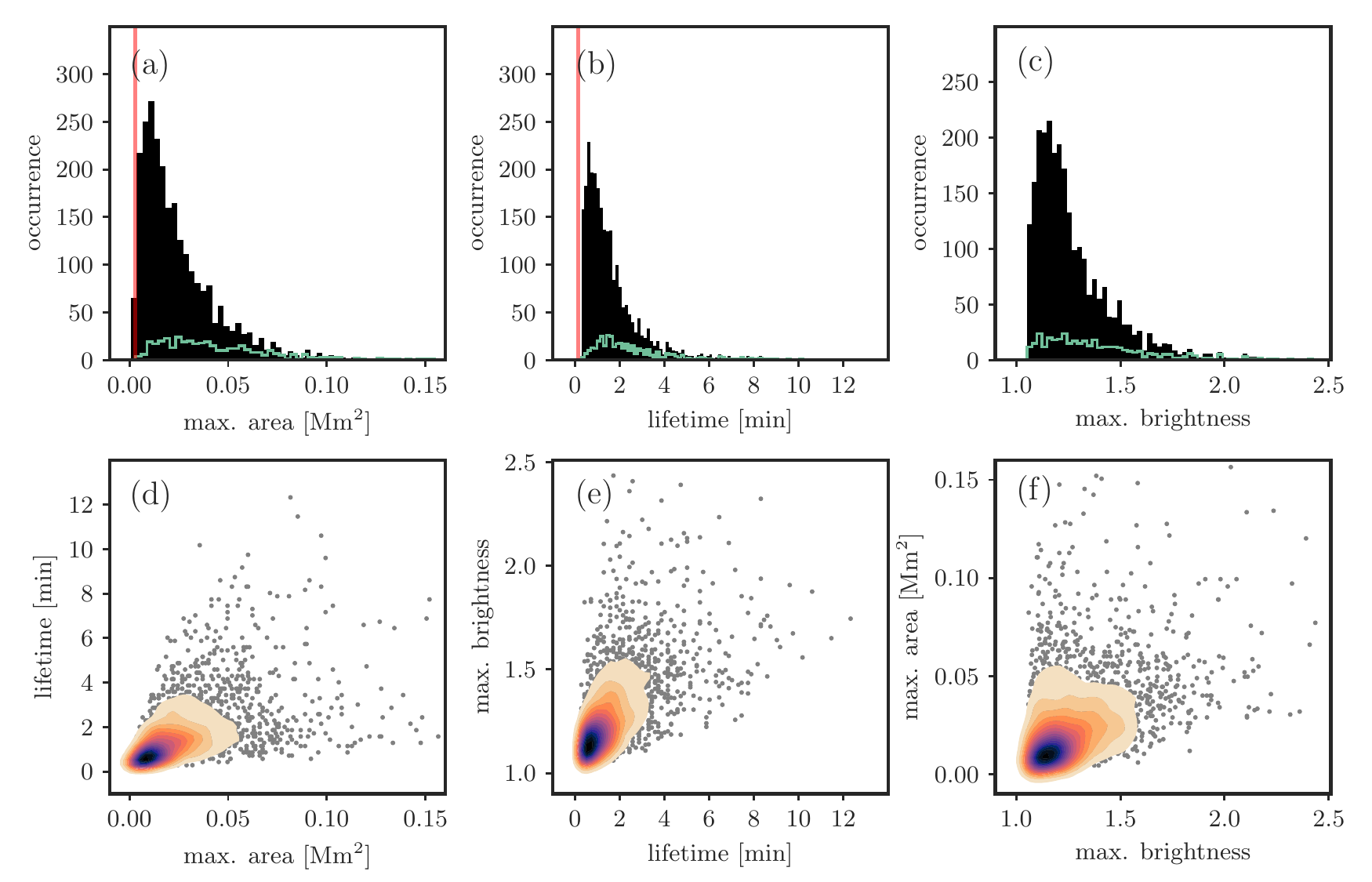}
\caption{\label{fig:stats}%
Maximum area, lifetime, and maximum brightness statistics of QSEBs. The total number of QSEBs is 2809. The filled black histograms in panels (a), (b), and (c) represent maximum area, lifetime, and maximum brightness distribution of QSEBs, respectively. Statistics of QSEBs with brightening in the \Hbeta{} line core (396 QSEBs) are presented by green histograms. The vertical red line in panels (a) and (b) mark the lower limit set by sampling: 0.0008~Mm$^2$ (1 pixel) in area and 8.6~s in lifetime. Panels (d--f): multivariate JPDFs and scatter plots between the maximum area, lifetime and, maximum brightness. The dark blue shade of JPDFs indicate highest density occurrence whereas the lighter orange shade regions represent low density distribution. 
One outlier QSEB that has the longest lifetime (20.5 min), largest maximum area (0.2603 Mm$^2$), and largest maximum brightness (2.76) is not shown in order to restrict the plotting ranges}.

\end{figure*}

\begin{figure}[!ht]
\includegraphics[width=0.49\textwidth]{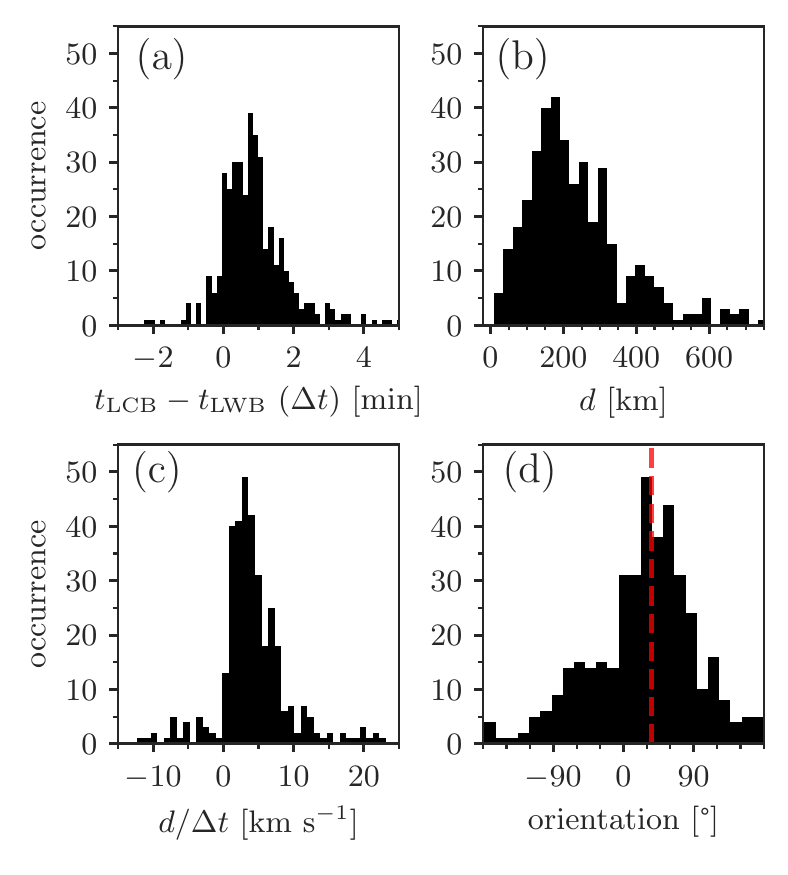}
\caption{\label{fig:lcb_stats}%
Time difference, distance, average propagation speed and orientation of brightening in the \Hbeta{} line core with respect to brightening in the wing. The method of measurement is illustrated in Fig.~\ref{fig:vel_est}. Positive values of propagation speed $d/\Delta t$ can be interpreted as upward propagation from lower to higher altitude. The vertical red dashed line in panel (d) indicates the direction towards the nearest limb (36\degr). The total number of measurements is 396.}
\end{figure}

\begin{figure*}[!ht]
\includegraphics[width=0.98\textwidth]{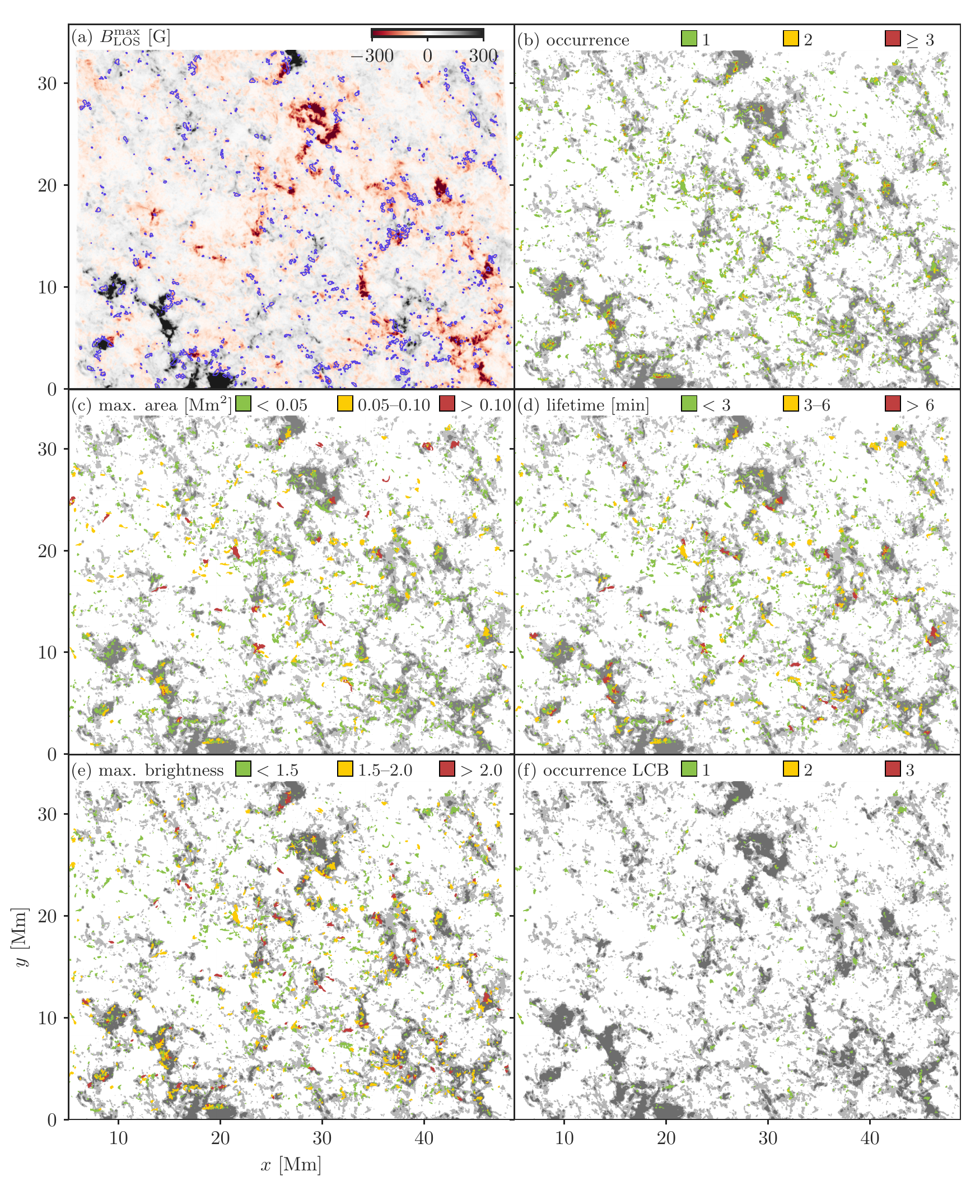}
\caption{\label{fig:spatial}%
Spatial distribution of QSEBs and their magnetic environment over the full 1~h duration of the time series. 
Panel (a) shows, at each pixel, 
the extremum of $B_{\rm{LOS}}$
over the whole 1~h duration of the time series. Blue contours mark pixels that have $\lvert B_{\rm{LOS}} \rvert > 50$~G for both polarities during the time series.
Panels (b) -- (e) display spatial location, area, lifetime and brightness of QSEBs, respectively. In case of multiple occurrence of QSEBs at a pixel (yellow and red areas in panel (b)), an average value of the parameters is presented in (c) -- (e). Panel (f) displays the location of QSEBs with line core brightening (LCB) in \Hbeta{}. The parameters presented in panels (b) -- (f) are segregated in three different bins with the values of the respective colors being given in the legend above each panel. The darker gray background in (b) -- (f) marks regions where $\lvert B_{\rm{LOS}}^{\rm{max}} \rvert$ is greater than 100~G, while lighter gray shade represents regions where $\lvert B_{\rm{LOS}}^{\rm{max}} \rvert$ is between 50--100~G.
}
\end{figure*}


\begin{figure*}[!ht]
\centering
\includegraphics[width=\textwidth]{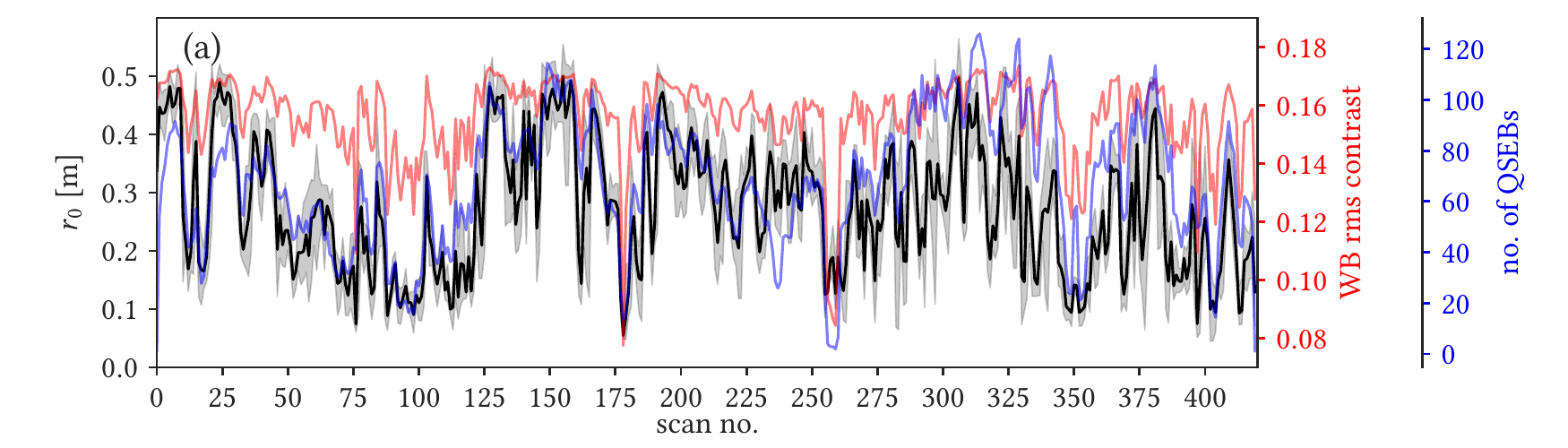}
\includegraphics[width=0.33\textwidth]{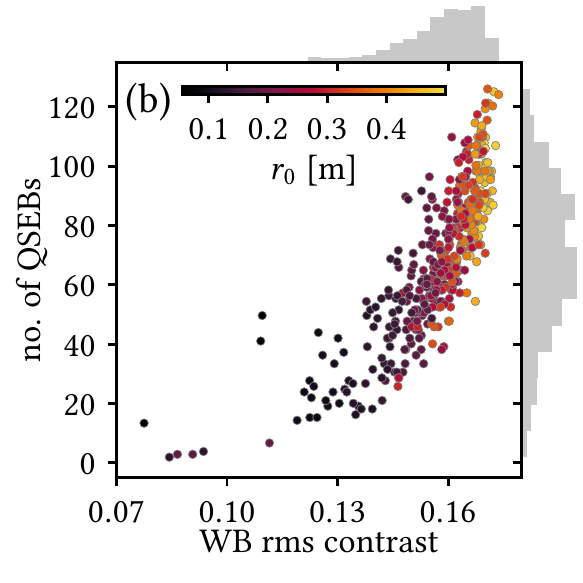}
\includegraphics[width=0.33\textwidth]{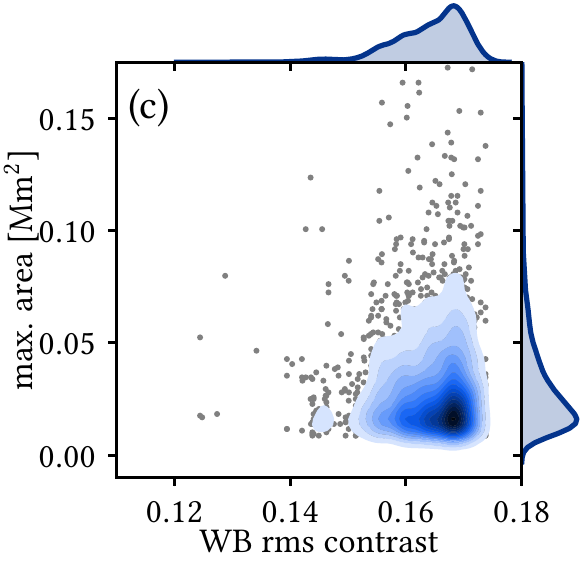}
\includegraphics[width=0.33\textwidth]{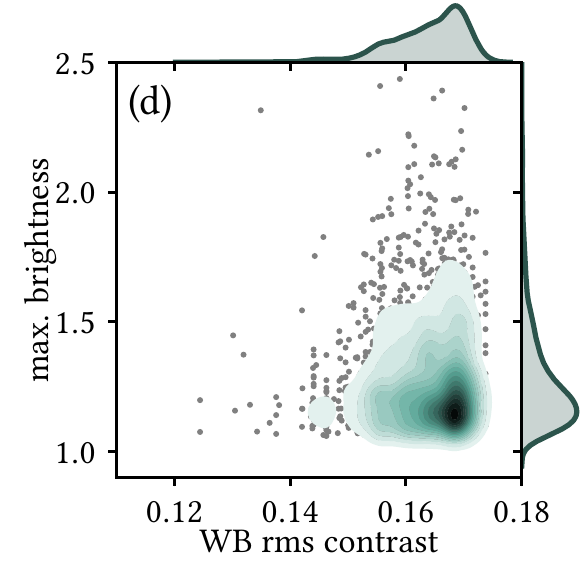}
\caption{\label{fig:WB_rms}%
Dependency of detected number of QSEBs and their measured area and brightness on atmospheric seeing conditions. (a) Variation of rms contrast of WB images (red), Fried's parameter $r_{\rm{0}}$ (black) and detected number of QSEBs (blue) during the observed time series. The shaded gray area indicates seeing ($r_{\rm{0}}$) variation within an \Hbeta{} line scan. (b) Scatter plot between rms contrast of WB images and no. of detected QSEBs. The colors of the data points show $r_{\rm{0}}$ values. (c) scatter and JPDFs between rms contrast of WB images and area of QSEBs. (d) scatter and JPDFs between rms contrast of WB images and brightness of QSEBs. Darker color shades in JPDFs plots indicate high density occurrence.  
}
\end{figure*}

The observations were obtained with the CHROMIS 
and CRISP
\citep{2008ApJ...689L..69S} 
imaging spectro(polari)meters at the Swedish 1-m Solar Telescope 
\citep[SST, ][]{2003SPIE.4853..341S} 
on 6 June 2019.
The target was a quiet-Sun region at $(x,y)=(611\arcsec,-7\arcsec)$ 
under a viewing angle $\mu=0.76$ (with $\mu=\cos \theta$ and $\theta$ the angle with the surface normal).
Figure~\ref{fig:overview} shows an overview of the observed field of view (FOV). 
The time series has a duration of 1~h and started at 8:41 UT. 

CHROMIS sampled the \Hbeta\ spectral line at 4861~\AA{} at 35 line positions between $\pm$1.371~\AA\ with 74~\mAA\ steps between $\pm$1.184~\AA. 
An \Hbeta{} line core image is shown in Fig.~\ref{fig:overview}b.
At each line position, a burst of 15 exposures was recorded which were used for image restoration. 
The exposure time was 8~ms and the time to complete a full \Hbeta\ spectral scan was 8.6~s.
CHROMIS has a transmission profile full-width at half-maximum (FWHM) of 100~\mAA\ at 4860~\AA, a pixel scale of 0\farcs038, and a FOV of 66\arcsec $\times$ 42\arcsec.
%
A sample WB image is shown in Fig.~\ref{fig:overview}a.
The CHROMIS instrument has an auxiliary wide-band (WB) channel that is equipped with a continuum filter centered at 4845~\AA\ (FWHM=6.5~\AA).
The WB channel serves as anchor channel for image restoration and is equipped with two cameras that are strictly synchronized with the CHROMIS narrow-band camera.
One of these cameras was put approximately 1 wave out of focus (3.35~mm) to allow for image restoration with phase diversity
\citep[following][]{2002SPIE.4792..146L}. 
The restored WB continuum images have the same cadence as the \Hbeta\ data.
More details on the optical setup on the SST imaging table are provided by 
\citet{2021A&A...653A..68L}. 

CRISP ran a program sampling the \Halpha, \ion{Fe}{i}~6173~\AA{}, and \ion{Ca}{ii}~8542~\AA{} spectral lines at a cadence of 35.9~s.
CRISP sampled the \Halpha{} line at 11 line positions between $\pm$1.5~\AA{} with 300~\mAA{} steps. 
Bursts of 6 exposures were acquired at each line position. 
The \ion{Fe}{i}~6173~\AA{} line was observed with polarimetry and was sampled at 13 line positions (between $\pm160$~\mAA\ with 40~\mAA\ steps, and further at $\pm240$~\mAA\ and $\pm320$~\mAA) plus the continuum at $+680$~\mAA\ from the nominal line core. 
Eight exposures per polarimetric state were acquired while the liquid crystal modulators were continuously cycling through four different states (this corresponds to 32 exposures per line position). 
Furthermore, spectropolarimetric observations were acquired in the \ion{Ca}{ii}~8542~\AA\ line in 20 line positions. 
The \Halpha{} observations were analyzed in \citetalias{2020A&A...641L...5J}, whereas the \ion{Ca}{ii}~8542~\AA\ data were not included in the analysis. 
%

High spatial resolution was achieved by the combination of good seeing conditions, the adaptive optics system and the high-quality CRISP and CHROMIS reimaging systems 
\citep{2019A&A...626A..55S}. 
We further applied image restoration using the multi-object multi-frame blind deconvolution 
\citep[MOMFBD, ][]{2005SoPh..228..191V} 
method. 
The data was processed with the standard SST data processing pipeline 
\citep{2015A&A...573A..40D, 
2021A&A...653A..68L}. 
The lower cadence and lower spatial resolution CRISP data (pixel scale 0\farcs058) were aligned to the CHROMIS data through cross-correlation of the WB channels that show similar photospheric scenes for both instruments. 
The CHROMIS field of view (FOV, approximately 66\arcsec $\times$ 45\arcsec) and temporal cadence served as reference to which the CRISP data was matched in space (FOV about 59\arcsec $\times$ 59\arcsec) by linear interpolation and in time by nearest-neighbor sampling. 
The alignment of the CRISP data included destretching to account for residual seeing-induced image deformation that was not accounted for by image restoration. 

We have performed Milne-Eddington inversions of the \ion{Fe}{i}~6173~\AA\ line data to infer the magnetic field vector 
utilizing a parallel C++/Python implementation\footnote{\url{https://github.com/jaimedelacruz/pyMilne}} \citep{2019A&A...631A.153D}.
A map of the line of sight magnetic field $B_\mathrm{LOS}$ is shown in Fig.~\ref{fig:overview}c.

\section{Methods and analysis}
\label{sec:method}
\subsection{Identification of QSEBs in \Hbeta{} spectra}
\label{sec:method_k-means}

We used the $k$-means clustering algorithm \citep{everitt_1972} 
to identify spectral signatures of QSEBs in \Hbeta{} spectra. 
The $k$-means method segregates $m$ number of data point with $n$ features into $k$ clusters.
In our case the data points are the spatially resolved image elements and features are the 35 wavelength positions sampled in the \Hbeta{} line. 
%
Each cluster is represented by a cluster center, i.e., the mean of all data points in a cluster. 
The clustering is improved through an iterative process for which the converging criterion is to minimize inertia, i.e., the within-cluster sum of squared euclidean distances from the cluster center. 
%
The algorithm is initialized by $k$ numbers of predefined centers and each data point is assigned to a closest (measured in euclidean distance) center thus creating the initial clusters. 
In each subsequent iteration, new centers are calculated from clusters defined in the previous iteration and this process continues until the algorithm converges. 
The $k$-means algorithm always converges but sometimes to a local minimum due to high dependency on the initialization, i.e., the initial selection for cluster centers. 
We used the $k$-means++ \citep[][]{arthur2007k} 
method for initialization which at first defines a cluster center from randomly selected data points and subsequently defines new cluster centers such that they are farthest from previously chosen centers.   

Before applying $k$-means clustering, it is important to determine the minimum number of clusters required to optimally represent the observations. 
We studied the change in the total inertia with respect to varying $k$ between 30 and 130
(a plot of the change in inertia is shown in Fig.~\ref{fig:no_cluster}).
Using the elbow method (see Appendix~\ref{app:kmeans}), we choose $k=100$.
%
%
A similar method was used by \citet{2019A&A...631L...5B} to determine the minimum number of clusters in an SST dataset in both \CaK{} and \Halpha{}.

We selected 40 scans with good seeing conditions out of the total of 420 scans spread over the whole time series to train our $k$-means model. 
The derived model was used to predict the closest cluster centers for each pixel in the complete time series. 
%
Hereafter, we refer to the cluster centers as representative profiles (RP). 
%
Figure~\ref{fig:overview}d shows the RP index map for one scan, demonstrating that each pixel belongs to one particular cluster. 
%
Out of the 100 RPs, we selected 24 RPs to detect QSEBs. These 24 selected clusters are shown in Fig.~\ref{fig:RPs}
(the remaining 76 RPs are shown in Fig.~\ref{fig:rest_RPs}).
Representative profiles 0--8 have the clearest characteristic spectral signature of QSEBs in the \Hbeta{} line, i.e., enhanced inner wings and unaffected line core. 
For RPs 9--15 only a weak intensity enhancement is found in the line wings.
It has been shown in \citetalias{2020A&A...641L...5J} that some QSEBs also exhibit brightening in the \Hbeta{} line core besides in the wings. 
Therefore, we also included RPs 16--23 which have elevated \Hbeta{} line core compare to the average profile.
The RPs with weak enhancement in the wings (RPs 9--15) and RPs with line core brightening (RPs 16--23) were only considered as a part of a QSEB if they appeared spatially and temporally in conjunction with RPs 0--8 which show the telltale sign of QSEBs; we elaborate more on this matter in the following sub-section. 

\subsection{Detection of QSEBs}
\label{sec:mothod_detection}

For the detection of QSEBs we located all the pixels which belong to one of the selected RPs.
For example, Fig.~\ref{fig:kmeans_label}(b) highlights all the pixels with selected RPs at one time step.  
We created binary images based on the spatial location of RPs 0--23, we refer to pixels with RPs 0--23 as ``foreground'' and the remaining pixels as ``background''. 
In order to track QSEBs in time, we performed a three-dimensional (3D) morphological closing operation to connect areas with selected RPs over multiple consecutive scans.
We used a $3\times3\times3$ structural element which covers the two spatial as well as the temporal dimension.
%
%
The 3D morphological operation for a QSEB is illustrated in ~Fig.~\ref{fig:evolution}. 
The closing operation fills gaps between the foreground pixels. 
For example, as can be seen in the third row of Fig.~\ref{fig:evolution}, the foreground pixels have gaps at time steps $t-\rm{t_{0}} =$ 103.2 and 129.0 sec, which are filled by the closing operations (see the bottom row).  
Similarly, any temporal gaps of one time step in the foreground pixels are also filled by the 3D closing operation. 
These temporal gaps are mostly caused by variations in image quality due to variable seeing conditions. 

We executed a 3D connected component labeling \citep{Fiorio_1996} on foreground pixels originating from the 3D morphological image processing.
The connected component labeling allows to uniquely label foreground pixels that are connected neighbors. 
The labeling algorithm requires a predefined criterion for connectivity. 
%
We prescribed a 26 neighborhood connectivity in 3D, i.e., two pixels are considered as connected if they share a face, or edge or corner. 
Through the described method we have detected 15938 ``events'', all uniquely labeled. 
For instance, the event shown in Fig.~\ref{fig:evolution} was labeled as event number 632. 
Moreover, not all the detected events are QSEBs. 
To qualify as QSEB, an event must have at least one pixel belonging to RPs 0--8 at any time during its lifetime.
A total of 2809 events satisfied the described condition, these were thus considered as QSEBs. 
%
Circles in the bottom panels of Fig.~\ref{fig:kmeans_label} mark a few example events that did not qualify as QSEB. 
We also excluded the events that have a lifetime shorter than two time steps (17.2~s) and have maximum area less than five pixels.
This means that a large event that appears in only one time step, or a small event that lives for more than two time steps, are still considered as valid QSEBs. 
In total we excluded 345 events that were too small and too short lived.       

%
Figure~\ref{fig:evolution} also explains the reason behind inclusion of RPs 9--15 in QSEB detection. 
At the onset, the QSEB has only weak intensity enhancement in the \Hbeta{} line wings and therefore at $t-\rm{t}_0=25.8$~s the QSEB pixels are identified by RPs 9--15. 
As the QSEB evolves, the central part exhibit higher intensity enhancement and is identified by RPs 0--8, however, pixels at the edges show weak intensity enhancement (RPs 9--15). 
Therefore, for an accurate measurement of lifetime and area of QSEBs, inclusion of RPs 9--15 is necessary. 

\subsection{Measuring properties of QSEBs}
\label{sec:mothod_properties}

In the next step we measured some basic properties like lifetime, maximum area, and maximum brightness of all the detected QSEBs. 
For the lifetime measurements we simply counted the number of scans from the start till the end of an event. 
For the area, we considered the scan when a QSEB occupied the maximum number of pixels.
%
The QSEB in Fig.~\ref{fig:evolution} had a lifetime of 137.6~s and covered an area of 0.0736~Mm$^2$ at the time of maximum area (at $t-\rm{t}_0=129$~s).

The maximum brightness of a QSEB was measured with respect to the averaged intensity in the local background. 
We located the pixel within a QSEB event that had maximum intensity enhancement in the line wings of the \Hbeta{} line. %
The obtained maximum intensity enhancement value was normalized to the far wing (average of the two extreme line positions sampled in the \Hbeta{} line, at $\Delta \lambda = \pm 1.371\AA$) intensity averaged over 50$\times$50  pixels surrounding the QSEB but excluding the QSEB pixels.   %
The QSEB in Fig.~\ref{fig:evolution} reached a maximum brightness of 2.54 (at $t-\rm{t}_0=120.4$~s). 

\paragraph{QSEBs with line core brightening.}

As mentioned earlier and reported in \citetalias{2020A&A...641L...5J}, some QSEBs exhibit brightening in the \Hbeta{} line core.
We have identified 396 QSEBs with line core brightening (14\% of the total number of detected QSEBs). 
In \citetalias{2020A&A...641L...5J} it is shown that the QSEB brightening in the \Hbeta{} line core appears with a temporal delay and spatial offset compared to the brightening in the line wings. 
We determined the temporal delay ($\Delta t$) and spatial offset ($d$) between line wing and line core brightening for all the QSEBs with line core brightening.      
For this purpose we considered the temporal difference between the first appearance of a QSEB in RPs 0--15 and the first appearance in RPs 16--23. 
For the spatial offset, we determined the separation between the centers of gravity of the area with line wing brightening and the area with line core brightening at their respective first appearances in a QSEB event. 
An example of this measurements is shown in Fig.~\ref{fig:vel_est}.
The orientation of the spatial offset between line wing brightening and line core brightening is measured with respect to horizontal direction as illustrated in Fig.~\ref{fig:vel_est}. 

\paragraph{Impact of seeing on QSEB detection.}
The image quality varies with atmospheric seeing conditions. 
In order to study the impact of seeing on the detection of QSEBs, we used measurements of the Fried's parameter $r_0$ and contrast variations in the WB images. 
The Fried's parameter is routinely measured at the SST from data taken with the wavefront sensor of the adaptive optics system 
\citep[see][]{2019A&A...626A..55S}. 
We used the measurements of $r_0$ that are mostly sensitive to near-ground seeing. 

The $r_0$ values varied between 4.5 and 56.5 cm, the WB contrast values varied between 7.8 and 17.4\%. 
Only 5 time steps had contrast values below 10\% and stand out for their poor image quality. 

\section{Results}
\label{sec:results}

\begin{table*}
\caption{Statistical properties of QSEBs}
\label{table:stats}      
\centering                                     
\begin{tabular}{c c c c c}          
\hline\hline                      
                   & \multicolumn{2}{c}{All QSEBs} & \multicolumn{2}{c}{QSEBs with LCB} \\
                        & mean   &  median & mean  & median \\ 
\hline                   
Max. area [Mm$^2$] (pixels)\tablefootmark{a}  & 0.0277 (36)  &  0.0203 (26)  &  0.0485 (62)  &  0.0396 (51)  \\     
Lifetime [min] (frames)\tablefootmark{a} & 1.65 (11)       &  1.14 (8)        &  2.63 (18)        &  2.00 (14) \\    
Max. brightness              & 1.28    &  1.22     &  1.39    &  1.33 \\     
\hline
\end{tabular}
\tablefoot{
\tablefoottext{a}{The values in parentheses are the nearest integer numbers.}
}
\end{table*}

Over the 1~h duration time series we detected a total of 2809 QSEBs.
Figure~\ref{fig:kmeans_label}a shows the detection of 91 QSEBs marked with red contours in a \Hbeta{} wing map recorded during excellent seeing ($r_0$ varying between 37 and 47~cm). 
The QSEBs appear typically as small and elongated brightenings in the \Hbeta{} wing images, we refer to \citetalias{2020A&A...641L...5J} 
where close-up images, as well as detailed spectral profiles and comparison with \Halpha{} for a number of examples are shown. 
The animation of Fig.~\ref{fig:kmeans_label} shows that the QSEBs are present all over the FOV. 

The $k$-means clustering method used in identifying QSEBs is able to distinguish between intensity enhancement associated with magnetic bright points and QSEBs. 
Magnetic bright points exhibit high intensity contrast in \Halpha{} and \Hbeta{} line wing images 
\citep{2006A&A...449.1209L} 
and can be easily mistaken for EB like events if EBs' detection is solely based on a contrast threshold applied to \Halpha{} or \Hbeta{} line wing images 
\citep{2013JPhCS.440a2007R}. 
As explained in Sect.~\ref{sec:method_k-means}, with the $k$-means clustering we select RPs which show intensity enhancement in the \Hbeta{} inner wings relative to the outer wings.      
%
Figure~\ref{fig:rest_RPs} shows that RPs 24--47 all have higher overall wing intensity than the average quiet Sun profile. These can be attributed to bright areas in granules and bright points.

The \Halpha{} and \Hbeta{} line wings forms under nearly local thermodynamical equilibrium conditions
\citep{2006A&A...449.1209L}, 
therefore, an enhancement in the inner wings compared to outer wings can be interpreted as a temperature enhancement in the upper photosphere relative to the atmosphere below. 
Figure~\ref{fig:kmeans_label} demonstrates the efficiency of our detection method in finding QSEBs and distinguishing them from magnetic bright points. 
For example, the bright points around $(x,y)=(30, 26)$~Mm in Fig~\ref{fig:kmeans_label}a are successfully eliminated through the detection procedure.    
Some QSEBs appear at and near bright points, in those situations our method only identifies part of a BP that shows the characteristic QSEB spectral signatures, for example, see bright points and QSEBs at $(\Delta x, \Delta y)=(1, 0)$~Mm in panel (e).

\subsection{Statistical properties}
\label{sec:stat_pro}
The distributions of the measured maximum area, lifetime and maximum brightness of all QSEBs are presented in Fig.~\ref{fig:stats}.
The maximum area varies between 0.0016~Mm$^2$ (2 pixels) and 0.2603 Mm$^2$ (338 pixels). 
We found QSEBs as short lived as 8.6~s (two time step) and 
the longest lived QSEB has a lifetime of 20.5~min (143 time steps). 
The maximum brightness of the QSEBs range between 1.06 and 2.76. 
The statistics shown in Fig.~\ref{fig:stats} exclude one outlier QSEB that has values for maximum area, maximum brightness, and lifetime that are far greater than for the other QSEBs.
All the distributions are positively skewed, i.e, the distributions have more weight towards the lower values and tail towards the higher values.   
The histogram of maximum area show a sharp cut-off at 0.0016~Mm$^2$ (two pixels).  
The mean and median values of the maximum area, lifetime, and maximum brightness is given in Table~\ref{table:stats}.
The joint probability distribution functions (JPDFs) and scatter plot between the three parameters are displayed in Fig.~\ref{fig:stats} with the purpose of analyzing their relationships. 
In general, 
there is a trend that QSEBs with bigger maximum area have longer lifetime and are also brighter. 
However, the scatter between these parameters is very large.
For example, several QSEBs that have smaller maximum area live long 
and some QSEBs with a short lifetime have big maximum area. 
%
%
Similar spread in relationships are present in the lifetime versus maximum brightness and maximum brightness versus maximum area JPDFs and scatter plots. 
\subsection{QSEBs with line core brightening}
\label{sec:stat_LCB}
For some QSEBs, brightening in the \Hbeta{} line wings also persist in the line core
\citepalias[see][]{2020A&A...641L...5J}.
The histograms of the maximum area, lifetime, and maximum brightness only for QSEBs with the line core brightening are also shown in Fig.~\ref{fig:stats}.
The mean and median values are given in Table~\ref{table:stats}. 
Qualitatively, these histograms do not stand out as different compared to those for all the QSEBs.
However, we found that 22.5\% of QSEBs with maximum area larger than 0.0203~Mm$^2$ (median value) exhibit brightening in the \Hbeta{} line core. 
Whereas, only 5.5\% QSEBs show the line core brightening if their maximum area is below 0.0203~Mm$^2$. 
It implies that, the bigger the maximum area, the higher the probability that a QSEB manifests the line core brightening.
Similar conclusions can be drawn about the lifetime and maximum brightness, i.e., the longer lived and brighter QSEBs are more likely to exhibit line core brightening. 

The QSEB examples shown in \citetalias{2020A&A...641L...5J} suggest that line core brightening appears with a temporal delay and spatial offset toward the nearest solar limb compared to its line wing counterparts. 
They interpret these results as upward propagation of reconnection brightening in vertically elongated current sheets and they found the propagation speed for these examples to vary between 3--10~\kms.
To put these results on a solid statistical footing, we analyzed all 396 QSEBs with line core brightening with the methods described in Section~\ref{sec:mothod_properties}. 
Figure~\ref{fig:lcb_stats}a shows the histogram of time difference ($\Delta t$) between line wings and line core brightening. 
A positive value of $\Delta t$ means that the brightening of line wings precede that of the line core.
We found only 27 (6.8\%) QSEBs with negative $\Delta t$, whereas 85.4\%  out of 396 QSEBs have $\Delta t$ between 0 and 3~min. 
The mean and median values for $\Delta t$ are 0.88~min and 0.72~min, respectively. 

The distance ($d$) between the areas of line core brightening and areas of line wings brightening ranges between 0 and 696~km (see Fig.~\ref{fig:lcb_stats}b). 
The mean and median values of $d$ are 238~km and 208~km, respectively. 
With the obtained values of $d$ and $\Delta t$ we measured the speed of propagation ($d/\Delta t$) of brightening from the line wings to line core. 
The mean and median values of $d/\Delta t$ are 4.4~\kms{} and 3.9~\kms{}, respectively.  
About 73\% of QSEBs with line core brightening have $d/\Delta t$ between 0 and 10~\kms, while the extreme values in the $d/\Delta t$ distribution are $-14.3$~\kms{} and 23.5~\kms. 

The distribution of the measured orientation of spatial offsets between line wings and line core brightening is presented in Fig.~\ref{fig:lcb_stats}d. 
The mean and median value of the orientation are 26.7\degr and 33.0\degr, respectively.   
The orientation of the direction to the limb closest to the center of the FOV is 36\degr. 
%
Most of the QSEBs exhibit line core brightening with a spatial offset towards the closest limb compared to line wings brightening. 
We found that 81.5\% have an orientation within $\pm90\degr$ from the closest limb direction (i.e., between $-54$\degr and $+126$\degr).
%
%

\subsection{Spatial distribution of QSEBs}
\label{sec:spatial}

The animation of Fig.~\ref{fig:kmeans_label} shows that QSEBs occur almost everywhere in the observed FOV. 
We analyzed the spatial distribution of QSEBs in detail with Fig.~\ref{fig:spatial}.
Panel (a) shows a map of the extreme values of $B_{\rm{LOS}}$ and blue contours mark areas where there have been significant magnetic fields ($\lvert B_{\rm{LOS}} \rvert > 50$~G) of both polarities. 
This illustrates that the occurrence of opposite polarities in close vicinity is very common, also in the network regions with strong magnetic field. 
For example, the positive polarity network patch at $(x,y)=(20,2)$~Mm is surrounded with blue contour patches at its outer perimeter. 
In Panel (b) we located all the pixels with QSEB occurrence during the 1~h long duration of the time series.
We also highlighted the pixels with multiple QSEB events with different colors.
%
The spatial distribution of QSEBs can be compared to the photospheric magnetic field that is shown as a background map. 
QSEBs can be found all over the FOV, however, QSEBs appear with higher temporal frequency at and close to the magnetic field concentrations in the network areas.
The inter-network is also covered by QSEB events but here we did not observe repetition of events (no yellow or red pixels, only green).   
The spatial distribution of QSEBs shows small voids which are approximately 3--6~Mm wide. 
In other words, there are finite empty spaces in the FOV where no QSEB events appear during 1~h. 
These voids are co-located with ``gaps'' in the $B_{\rm{LOS}}$ maps where $\lvert B_{\rm{LOS}} \rvert<50$~G.
%

%
%
Panels (c) -- (e) address the question whether there is a spatial correlation with respect to the maximum area, lifetime, and maximum brightness. 
The QSEBs with larger maximum area (>0.10~Mm$^2$) shown by red color in panel (c) predominantly occur at and close to larger and stronger magnetic field concentrations in the network regions.  
On the other hand, QSEBs with smaller (<0.05~Mm$^2$) and intermediate (0.05--0.10 Mm$^2$) maximum area do not have any spatial preference and appear both in the network and inter-network areas.     
Similar behaviors are found for QSEBs which have longer lifetime (>6~min) or larger brightness (>2.0) or both, i.e., these QSEBs largely take place in the network regions. 
Moreover, shorter lived and less brighter QSEBs do not exhibit any spatial preference. 
%

%
%
%
Panel (f) of Fig.~\ref{fig:spatial} shows the spatial distribution of QSEBs with line core brightening. 
There appears to be no spatial preference for these QSEBs and they are nearly uniformly distributed over the FOV. 

\subsection{Impact of atmospheric seeing  on QSEB detection}
\label{sec:seeing_impact}

Even during the most favorable weather conditions, ground based solar observations are prone to variable atmospheric seeing.  
We detected a significant number of QSEBs whose maximum area and life time are close to the spatial and temporal resolution limit. 
Therefore, we analyzed the impact of variations in atmospheric seeing on the detection of QSEBs.
%
%
The detected number of QSEBs in each \Hbeta{} scan are shown and compared with the seeing condition as measured by the Fried's parameter $r_{\rm{0}}$ and WB image contrast in Fig.~\ref{fig:WB_rms}.
It is evident that the number of detected QSEBs in a \Hbeta{} scan highly depends on the $r_{\rm{0}}$ values at the time when the scan is recorded. 
The scatter plot between WB image contrast and number of QSEBs indicate that we detect higher number of QSEBs in higher quality images.
%
Even a slight variation in the
image quality and seeing conditions severely affects the detection of QSEBs.
Our best quality WB images have around 17\% contrast ($r_0\gtrsim40$~cm) and in those scans, on average we detected about 100 QSEBs.
%
In the scans where WB images have around 16\% contrast, on average, we found only 65 QSEBs. 
We see further reduction in the number of detected QSEBs with further decrease in contrast of the WB images.  

The measurements of QSEB properties like maximum area and brightness are also affected by the variations in atmospheric seeing. 
Figure~\ref{fig:WB_rms}{c} shows the scatter plot and JPDFs between contrast of WB images and maximum area of QSEBs. 
During the best seeing conditions where WB images have contrast above 16\%, we detected QSEBs with maximum area ranging between 0.0016~Mm$^2$ and 0.18~Mm$^2$. 
As the contrast of WB images decreases, the measured maximum area of QSEBs tends to be smaller.
Similarly, when the contrast of WB images is greater than 16\%, we observed QSEBs with the maximum brightness reaching up to 2.4 (see panel~(d)). 
However, as the seeing degrades the measured maximum brightness of QSEBs is restricted to lower values. 
For QSEBs observed with WB image contrast below 15\%, 
the measured maximum brightness is below 1.8, with one outlier QSEB having a maximum brightness of 2.3.

Similar to the maximum area and maximum brightness, the measured lifetimes of the QSEBs can be affected by the varying seeing conditions. Due to a drop in the seeing conditions, a QSEB can appear with a delay, disappear prematurely or disappear and reappear again, thus affecting the lifetime measurement.
Since the seeing conditions were consistently of high quality, with only a few interruptions of very bad quality, we regard the measurements of long duration QSEBs as reliable. It is more likely that we underestimate the number of short-duration QSEBs due to rapid seeing variations. 


%


\section{Discussions and conclusions}
\label{sec:conclusion}


%
%

%

We performed a detailed statistical analysis of small-scale magnetic reconnection events in the lower solar atmosphere which were recently reported to be ubiquitous in the quiet Sun \citepalias{2020A&A...641L...5J}. 
Using $k$-means clustering followed by morphological operations we detected a total of 2809 QSEBs over a FOV of 47 $\times$ 32~Mm and a duration of 1~hr.   
%
We performed an extensive statistical characterization of these QSEBs and measured lifetimes, maximum area, maximum brightness and the spatial distribution over the FOV.

The maximum area occupied by the QSEBs during their lifetime varied between 0.0016~Mm$^2$ and 0.2630~Mm$^2$. 
The distribution of QSEB maximum area is positively skewed, i.e., QSEBs with smaller maximum area were observed with higher frequency, 
while fewer and fewer QSEBs were found with increasing maximum area. 
Towards smaller scales, the maximum area distribution has a sharp cut off near the spatial resolution limit indicating that a significant number of QSEBs were not fully resolved in the presented observations. 
Therefore, observations at even higher spatial resolution 
(i.e., better than 0\farcs1)
are pivotal to fully explore the properties of QSEBs.  

The QSEB lifetime were found to range between 8.6~s and 20.5~min. 
We found higher numbers of QSEBs with shorter lifetime (<2~min) and relatively fewer QSEBs with longer lifetime.
The median QSEB lifetime was 1.14~min.    
We found 48 QSEB events which appear only in two time steps, suggesting that these events are not temporally resolved. 
The maximum brightness of QSEBs have a distribution similar to those for maximum area and lifetime, i.e., postively skewed with more QSEBs with weaker brightness enhancement and fewer with strong brightness enhancement. 

The JPDFs and scatter plot between maximum area, lifetime, and maximum brightness indicate that 
roughly speaking, QSEBs with bigger maximum area also have longer lifetime and higher brightness enhancement. 
However, there is a large spread and we also observed ample events which had a bigger maximum area but lived for a shorter duration and exhibited a weaker maximum brightening. 
Similarly, there were QSEBs events that had a smaller maximum area but were long lived and had a higher maximum brightness.  

\citet{2019A&A...626A...4V} 
analyzed the lifetime, area, and brightness contrast of 1735 EBs from 10 different active regions observed in the \Halpha{} line with SST. 
They found that the median value of EB lifetimes was about 3~min which is approximately 2.5 times longer than for the QSEB lifetimes determined in this paper. 
Their results also suggest that EBs are approximately four times bigger in area than QSEBs. 
The median value of area of the EBs observed by 
\citet{2019A&A...626A...4V} 
was 0.076~Mm$^2$ (0\farcs14$^2$), whereas we found that the median value of QSEB maximum area is 0.0203~Mm$^2$.
On average, the EBs observed by 
\citet{2019A&A...626A...4V} 
had higher brightness compared to the QSEBs reported in this paper. 
Distributions of maximum area, lifetime and brightness of active region EBs observed by these authors are also positively skewed and are qualitatively very similar to those of QSEBs presented in our analysis. 
\citet{2019A&A...626A...4V} report an occurrence rate of 1.1~arcmin$^{-2}\,$min$^{-1}$ (5.7$\times10^{-4}$~Mm$^{-2}\,$min$^{-1}$) for EBs, which is at least an order of magnitude lower than the occurrence rate of 60.8~arcmin$^{-2}\,$min$^{-1}$ (3.1$\times10^{-2}$~Mm$^{-2}\,$min$^{-1}$) found for QSEBs in our analysis. 
We note that 
\citet{2019A&A...626A...4V} 
observed the EBs in the \Halpha{} line, whereas QSEBs observations reported in this paper are observed in the \Hbeta{} line. 
Due to the shorter wavelength, \Hbeta{} observations provide better spatial resolution and higher temperature sensitivity, and thus are more effective in detection of weaker and smaller QSEB and EB events. 
Therefore, a comprehensive comparison of EB and QSEB properties requires observations of both phenomena in the \Hbeta{} line. 
We anticipate that active region observations in the \Hbeta{} line will reveal higher occurrence of EBs.  
For a literature review on statistical properties of EBs we refer to   
\citet{2019A&A...626A...4V}.  
\citet{2021A&A...648A..54R} 
observed EBs in the sunspot moat region and penumbra  (penumbral EBs, or PEBs) in the \Hbeta{} line with SST. 
Their results show that the EBs in the moat region have number density of 1.72~Mm$^{-2}$ and PEBs have number density of 0.76~Mm$^{-2}$. 
This is a factor 19.1 and 8.4 times higher than the number density of QSEBs (0.09~Mm$^{-2}$) we found here. 
%

%
The characteristic hydrogen Balmer spectral signature of EBs is an intensity enhancement in the line wings with unaffected line core.
For the majority of the observed QSEBs, we found such spectral signatures in the \Hbeta{} line. 
However, 14\% of QSEBs manifest compact brightening in the \Hbeta{} line core in tandem with their line wings counterparts.  
Moreover, line core brightenings exhibit a spatial and temporal offset 
with respect to line wings brightenings. 
In around 93\% of the events, the line core brightening occurs with a time delay with respect to the onset of brightening in the line wings. 
The median value of the temporal delay is 53~s, while the median value of spatial offset between areas of line wings and line core brightenings was found to be 204~km. 
In the majority of events we found that the spatial offsets in the line core and line wings brightening locations are oriented in and close to the direction of the closest limb with line core brightening appearing relatively closer to the limb. 
Since the observed FOV was away from the disk center ($\mu=0.76$), the QSEBs were viewed from the side under inclined angle, the temporal delay and spatial offset in line core brightenings can be interpreted as upward propagating brightening from the photosphere towards the lower chromopshere in vertically elongated current sheets.   
Our measurements suggests that the reconnection brightening in QSEBs propagates upward with speeds ranging between 0 and 23~\kms. 
QSEBs with bigger area, longer lifetime and higher brightness have higher probability to exhibit line core brightening in the \Hbeta{} line. 
%
As discussed in \citetalias{2020A&A...641L...5J}, the observation of \Hbeta\ line core brightening and propagtion of the brightening aligns well with the vertical current sheets in the simulations of \citet{2019A&A...626A..33H}. 
These simulations demonstrate the occurrence of EBs and UV bursts
\citep{2014Sci...346C.315P, 
2018SSRv..214..120Y} 
along extended current sheets with EBs located in the deeper part of the atmosphere and UV bursts in the higher atmosphere.
A spatial offset between EBs and UV bursts in off-center observations were observed by
\citet{2015ApJ...812...11V} 
and 
\citet{2019ApJ...875L..30C}. 
The observation of transition region \ion{Si}{iv} emission associated with 2 QSEBs by \citet{2017ApJ...845...16N} 
is also consistent with a scenario of reconnection along a vertical current sheet in QSEBs.

We found that QSEBs are nearly uniformly distributed over the observed FOV, i.e., they occur everywhere in the quiet Sun including the network and inter-network regions. 
However, in the network regions, QSEBs appear more frequently. 
Repetitions of QSEB events at one particular location in the inter-network region are rarely observed during the 1~h long time-series. 
We observed that bigger, longer lived, and brighter QSEBs occur in the vicinity of the network magnetic field concentrations. 
Whereas, QSEBs with smaller to intermediate maximum area, lifetime, and maximum brightness occur everywhere in the FOV.   
The differences between the properties of QSEBs appearing in the network and inter-network regions could be explained by the disparity of magnetic flux and energy between these regions. 
Similar interpretation holds for active region EBs being bigger, long lived and brighter compared to QSEBs.
Our results indicate that the QSEBs with line core brightening in the \Hbeta{} line do not have any spatial preference and appear evenly over the FOV.  

Even though the QSEBs were ubiquitous and nearly uniformly distributed in the FOV, we found small voids of 3--6~Mm wide in the inter-network regions.
In these voids no QSEB events occurred during our 1~h long observations.
The voids in the spatial distribution of QSEBs are coinciding with areas where the magnetic field remained very weak ($\lvert B_{\rm{LOS}}\rvert <$ 50~G) throughout the observations. 
The spatial scale of these voids is similar to the spatial scale of mesogranulation \citep[see][]{1990ARA&A..28..263S}. 
The granulation and supergranulation are two distinctively recognizable convective patterns observed in the photosphere. 
However, existence of the mesogranular convective scale is still under debate.  
The mesogranulation patterns which have average diameter of 5~Mm and lifetime of 3~h are mostly seen in horizontal flow divergence maps derived by tracking granules
\citep[see, for example,][among others]{1988ApJ...333..427N,1992Natur.356..322M, 
2005A&A...444..245L}. 
On the other hand, the Fourier power spectra of photospheric Doppler maps do not reveal any distinct convective scale corresponding to mesogranulation 
\citep[see, for example,][among others]{1989SoPh..123...21W, 
2012ApJ...758..139K}.   
\citet{2011ApJ...727L..30Y} confirmed the absence of a discrete convective scale of mesogranular size, without denying its presence as a part of the convective power spectrum. 
However, these authors showed that 80\% of magnetic elements with flux density above 30~G are concentrated in and around mesogranular lanes. 
%
%
Our analysis of the $B_{\rm{LOS}}$ maps (see Fig.~\ref{fig:spatial}a) is in agreement with the results of \citet{2011ApJ...727L..30Y} in that we found regions with only weak fields ($\lvert B_{\rm{LOS}}\rvert <$ 50~G) with a typical mesogranular size. 
QSEBs occur at the edges of these regions and suggest that they require magnetic fields that are stronger than the weakest fields in the quiet Sun.

While the observing angle for this dataset is advantageous for viewing the characteristic EB flame morphology, it is not optimal for studying the detailed relationship between the photospheric magnetic field topology and QSEB occurence. 
We find that we do not always have a clear view on magnetic fields in intergranular lanes as they are sometimes hidden behind granular ``hill'' tops in the foreground. %
A study that can unambiguously track magnetic fields rooted in the photosphere and their relation to QSEB occurrence and evolution requires time sequences of observations more closer to the disk center that have unobstructed view on the intergranular lanes.

In the best quality scan we found 126 QSEBs in the FOV. 
Our rough extrapolation suggests that there could be as many as half a million QSEBs present on the solar surface at any given time. 
%
With this estimate, we neglect the possibility of regional variations in QSEB population due to differences in magnetic activity and topology.
For example, the QSEB population density could be different in enhanced network regions and in coronal holes.

%
QSEBs are difficult to observe because of their sub-arcsec spatial size and 
limited brightness enhancement and therefore require excellent quality observations.
Our analysis on the efficiency in the detection of QSEBs in relationship with seeing variations clearly shows that even a slight change in seeing conditions severely affects the detection of QSEBs. 
We detected up to 126 QSEBs in our best quality \Hbeta{} scans which have rms contrast of 17\%. 
%
On the other hand, with 16\%  WB rms contrast we found only 69 QSEBs on average, that is 45\% reduction in detected QSEBs with only 1\% reduction in the WB rms contrast.   
If we assume that on average 100 QSEBs should be present in the FOV all the time 
and take a typical lifetime of 1.14~min, 
we estimate that under continuous excellent seeing conditions, we could have detected a total of 5250 QSEBs.
This is 1.8 times higher than the actual detected number of QSEBs.
We detected 12157 events which exhibited only very weak intensity enhancement in the \Hbeta{} line wings (clustered as RPs 9--15, see Fig.~\ref{fig:RPs}) and were not considered as QSEB. 
Some of these events could be actual QSEBs which remained undetected due an inadequate image contrast caused by the seeing variations. 
Furthermore, we observed significant number of QSEBs which have maximum area close to the spatial resolution limit.
Therefore, we anticipate that a fraction of these events with weak intensity enhancement could also be actual QSEBs, but not fully resolved due to the spatial resolution limit (0\farcs1).

We conclude that QSEBs are present in large numbers in quiet Sun and appear everywhere except in areas of mesogranular size with weakest magnetic field.
Given the high number density, follow-up studies on their impact on the lower solar atmosphere and establish the role of QSEBs in the mass and energy transfer in the solar atmopshere are waranted. 
We show that a spatial resolution better than 0\farcs1 is required and this makes QSEBs an excellent target for 
the 4-m DKIST telescope \citep{2020SoPh..295..172R} 
and the planned
EST \citep{2019arXiv191208650S}. 
The QSEB phenomenon provides a view on the fundamental process of magnetic reconnection on the smallest spatial scales observable in astrophysics. 

\begin{acknowledgements}
The Swedish 1-m Solar Telescope is operated on the island of La Palma
by the Institute for Solar Physics of Stockholm University in the
Spanish Observatorio del Roque de los Muchachos of the Instituto de
Astrof{\'\i}sica de Canarias.
The Institute for Solar Physics is supported by a grant for research infrastructures of national importance from the Swedish Research Council (registration number 2017-00625).
This research is supported by the Research Council of Norway, project numbers 250810, 
325491, 
and through its Centres of Excellence scheme, project number 262622.
We made much use of NASA's Astrophysics Data System Bibliographic Services.
\end{acknowledgements}

\bibliographystyle{aa-note}
\bibliography{QSEB_pro} 

\begin{thebibliography}{52}
\expandafter\ifx\csname natexlab\endcsname\relax\def\natexlab#1{#1}\fi

\bibitem[{Arthur \& Vassilvitskii(2007)}]{arthur2007k}
Arthur, D. \& Vassilvitskii, S. 2007, in Proceedings of the eighteenth annual
  ACM-SIAM symposium on Discrete algorithms, Society for Industrial and Applied
  Mathematics, 1027--1035 \csname arthur2007klink\endcsname~\csname
  arthur2007knote\endcsname

\bibitem[{{Bose} {et~al.}(2019){Bose}, {Henriques}, {Joshi}, \& {Rouppe van der
  Voort}}]{2019A&A...631L...5B}
{Bose}, S., {Henriques}, V. M.~J., {Joshi}, J., \& {Rouppe van der Voort}, L.
  2019, \aap, 631, L5 \csname 2019A&A...631L...5Blink\endcsname~\csname
  2019A&A...631L...5Bnote\endcsname

\bibitem[{{Chen} {et~al.}(2019){Chen}, {Tian}, {Peter}, {Samanta},
  {Yurchyshyn}, {Wang}, {Cao}, {Wang}, \& {He}}]{2019ApJ...875L..30C}
{Chen}, Y., {Tian}, H., {Peter}, H., {et~al.} 2019, \apjl, 875, L30 \csname
  2019ApJ...875L..30Clink\endcsname~\csname 2019ApJ...875L..30Cnote\endcsname

\bibitem[{{da Silva Santos} {et~al.}(2020){da Silva Santos}, {de la Cruz
  Rodr{\'\i}guez}, {White}, {Leenaarts}, {Vissers}, \&
  {Hansteen}}]{2020A&A...643A..41D}
{da Silva Santos}, J.~M., {de la Cruz Rodr{\'\i}guez}, J., {White}, S.~M.,
  {et~al.} 2020, \aap, 643, A41 \csname
  2020A&A...643A..41Dlink\endcsname~\csname 2020A&A...643A..41Dnote\endcsname

\bibitem[{{Danilovic}(2017)}]{2017A&A...601A.122D}
{Danilovic}, S. 2017, \aap, 601, A122 \csname
  2017A&A...601A.122Dlink\endcsname~\csname 2017A&A...601A.122Dnote\endcsname

\bibitem[{{de la Cruz Rodr{\'\i}guez}(2019)}]{2019A&A...631A.153D}
{de la Cruz Rodr{\'\i}guez}, J. 2019, \aap, 631, A153 \csname
  2019A&A...631A.153Dlink\endcsname~\csname 2019A&A...631A.153Dnote\endcsname

\bibitem[{{de la Cruz Rodr{\'{\i}}guez} {et~al.}(2015){de la Cruz
  Rodr{\'{\i}}guez}, {L{\"o}fdahl}, {S{\"u}tterlin}, {Hillberg}, \& {Rouppe van
  der Voort}}]{2015A&A...573A..40D}
{de la Cruz Rodr{\'{\i}}guez}, J., {L{\"o}fdahl}, M.~G., {S{\"u}tterlin}, P.,
  {Hillberg}, T., \& {Rouppe van der Voort}, L. 2015, \aap, 573, A40 \csname
  2015A&A...573A..40Dlink\endcsname~\csname 2015A&A...573A..40Dnote\endcsname

\bibitem[{{Ellerman}(1917)}]{1917ApJ....46..298E}
{Ellerman}, F. 1917, \apj, 46, 298 \csname
  1917ApJ....46..298Elink\endcsname~\csname 1917ApJ....46..298Enote\endcsname

\bibitem[{Everitt(1972)}]{everitt_1972}
Everitt, B.~S. 1972, British Journal of Psychiatry, 120, 143–145 \csname
  everitt_1972link\endcsname~\csname everitt_1972note\endcsname

\bibitem[{{Fang} {et~al.}(2006){Fang}, {Tang}, {Xu}, {Ding}, \&
  {Chen}}]{2006ApJ...643.1325F}
{Fang}, C., {Tang}, Y.~H., {Xu}, Z., {Ding}, M.~D., \& {Chen}, P.~F. 2006,
  \apj, 643, 1325 \csname 2006ApJ...643.1325Flink\endcsname~\csname
  2006ApJ...643.1325Fnote\endcsname

\bibitem[{Fiorio \& Gustedt(1996)}]{Fiorio_1996}
Fiorio, C. \& Gustedt, J. 1996, Theoretical Computer Science, 154, 165  \csname
  Fiorio_1996link\endcsname~\csname Fiorio_1996note\endcsname

\bibitem[{{Georgoulis} {et~al.}(2002){Georgoulis}, {Rust}, {Bernasconi}, \&
  {Schmieder}}]{2002ApJ...575..506G}
{Georgoulis}, M.~K., {Rust}, D.~M., {Bernasconi}, P.~N., \& {Schmieder}, B.
  2002, \apj, 575, 506 \csname 2002ApJ...575..506Glink\endcsname~\csname
  2002ApJ...575..506Gnote\endcsname

\bibitem[{{Hansteen} {et~al.}(2019){Hansteen}, {Ortiz}, {Archontis},
  {Carlsson}, {Pereira}, \& {Bj{\o}rgen}}]{2019A&A...626A..33H}
{Hansteen}, V., {Ortiz}, A., {Archontis}, V., {et~al.} 2019, \aap, 626, A33
  \csname 2019A&A...626A..33Hlink\endcsname~\csname
  2019A&A...626A..33Hnote\endcsname

\bibitem[{{Hansteen} {et~al.}(2017){Hansteen}, {Archontis}, {Pereira},
  {Carlsson}, {Rouppe van der Voort}, \& {Leenaarts}}]{2017ApJ...839...22H}
{Hansteen}, V.~H., {Archontis}, V., {Pereira}, T.~M.~D., {et~al.} 2017, \apj,
  839, 22 \csname 2017ApJ...839...22Hlink\endcsname~\csname
  2017ApJ...839...22Hnote\endcsname

\bibitem[{{Joshi} {et~al.}(2020){Joshi}, {Rouppe van der Voort}, \& {de la Cruz
  Rodr{\'{\i}}guez}}]{2020A&A...641L...5J}
{Joshi}, J., {Rouppe van der Voort}, L. H.~M., \& {de la Cruz
  Rodr{\'{\i}}guez}, J. 2020, \aap, 641, L5 \csname
  2020A&A...641L...5Jlink\endcsname~\csname 2020A&A...641L...5Jnote\endcsname

\bibitem[{{Katsukawa} \& {Orozco Su{\'a}rez}(2012)}]{2012ApJ...758..139K}
{Katsukawa}, Y. \& {Orozco Su{\'a}rez}, D. 2012, \apj, 758, 139 \csname
  2012ApJ...758..139Klink\endcsname~\csname 2012ApJ...758..139Knote\endcsname

\bibitem[{{Kurokawa} {et~al.}(1982){Kurokawa}, {Kawaguchi}, {Funakoshi}, \&
  {Nakai}}]{1982SoPh...79...77K}
{Kurokawa}, H., {Kawaguchi}, I., {Funakoshi}, Y., \& {Nakai}, Y. 1982,
  \solphys, 79, 77 \csname 1982SoPh...79...77Klink\endcsname~\csname
  1982SoPh...79...77Knote\endcsname

\bibitem[{{Leenaarts} {et~al.}(2006){Leenaarts}, {Rutten}, {S{\"u}tterlin},
  {Carlsson}, \& {Uitenbroek}}]{2006A&A...449.1209L}
{Leenaarts}, J., {Rutten}, R.~J., {S{\"u}tterlin}, P., {Carlsson}, M., \&
  {Uitenbroek}, H. 2006, \aap, 449, 1209 \csname
  2006A&A...449.1209Llink\endcsname~\csname 2006A&A...449.1209Lnote\endcsname

\bibitem[{{Leitzinger} {et~al.}(2005){Leitzinger}, {Brandt}, {Hanslmeier},
  {P{\"o}tzi}, \& {Hirzberger}}]{2005A&A...444..245L}
{Leitzinger}, M., {Brandt}, P.~N., {Hanslmeier}, A., {P{\"o}tzi}, W., \&
  {Hirzberger}, J. 2005, \aap, 444, 245 \csname
  2005A&A...444..245Llink\endcsname~\csname 2005A&A...444..245Lnote\endcsname

\bibitem[{{Libbrecht} {et~al.}(2017){Libbrecht}, {Joshi}, {Rodr{\'\i}guez},
  {Leenaarts}, \& {Ramos}}]{2017A&A...598A..33L}
{Libbrecht}, T., {Joshi}, J., {Rodr{\'\i}guez}, J. d. l.~C., {Leenaarts}, J.,
  \& {Ramos}, A.~A. 2017, \aap, 598, A33 \csname
  2017A&A...598A..33Llink\endcsname~\csname 2017A&A...598A..33Lnote\endcsname

\bibitem[{{L{\"o}fdahl}(2002)}]{2002SPIE.4792..146L}
{L{\"o}fdahl}, M.~G. 2002, in Presented at the Society of Photo-Optical
  Instrumentation Engineers (SPIE) Conference, Vol. 4792, Society of
  Photo-Optical Instrumentation Engineers (SPIE) Conference Series, ed.
  {P.~J.~Bones, M.~A.~Fiddy, \& R.~P.~Millane}, 146--155 \csname
  2002SPIE.4792..146Llink\endcsname~\csname 2002SPIE.4792..146Lnote\endcsname

\bibitem[{{L{\"o}fdahl} {et~al.}(2021){L{\"o}fdahl}, {Hillberg}, {de la Cruz
  Rodr{\'\i}guez}, {Vissers}, {Andriienko}, {Scharmer}, {Haugan}, \&
  {Fredvik}}]{2021A&A...653A..68L}
{L{\"o}fdahl}, M.~G., {Hillberg}, T., {de la Cruz Rodr{\'\i}guez}, J., {et~al.}
  2021, \aap, 653, A68 \csname 2021A&A...653A..68Llink\endcsname~\csname
  2021A&A...653A..68Lnote\endcsname

\bibitem[{{Matsumoto} {et~al.}(2008){Matsumoto}, {Kitai}, {Shibata}, {Nagata},
  {Otsuji}, {Nakamura}, {Watanabe}, {Tsuneta}, {Suematsu}, {Ichimoto},
  {Shimizu}, {Katsukawa}, {Tarbell}, {Lites}, {Shine}, \&
  {Title}}]{2008PASJ...60..577M}
{Matsumoto}, T., {Kitai}, R., {Shibata}, K., {et~al.} 2008, \pasj, 60, 577
  \csname 2008PASJ...60..577Mlink\endcsname~\csname
  2008PASJ...60..577Mnote\endcsname

\bibitem[{{Muller} {et~al.}(1992){Muller}, {Auffret}, {Roudier}, {Vigneau},
  {Simon}, {Frank}, {Shine}, \& {Title}}]{1992Natur.356..322M}
{Muller}, R., {Auffret}, H., {Roudier}, T., {et~al.} 1992, \nat, 356, 322
  \csname 1992Natur.356..322Mlink\endcsname~\csname
  1992Natur.356..322Mnote\endcsname

\bibitem[{{Nelson} {et~al.}(2017){Nelson}, {Freij}, {Reid}, {Oliver},
  {Mathioudakis}, \& {Erd{\'e}lyi}}]{2017ApJ...845...16N}
{Nelson}, C.~J., {Freij}, N., {Reid}, A., {et~al.} 2017, \apj, 845, 16 \csname
  2017ApJ...845...16Nlink\endcsname~\csname 2017ApJ...845...16Nnote\endcsname

\bibitem[{{Nelson} {et~al.}(2015){Nelson}, {Scullion}, {Doyle}, {Freij}, \&
  {Erd{\'e}lyi}}]{2015ApJ...798...19N}
{Nelson}, C.~J., {Scullion}, E.~M., {Doyle}, J.~G., {Freij}, N., \&
  {Erd{\'e}lyi}, R. 2015, \apj, 798, 19 \csname
  2015ApJ...798...19Nlink\endcsname~\csname 2015ApJ...798...19Nnote\endcsname

\bibitem[{{Nelson} {et~al.}(2013){Nelson}, {Shelyag}, {Mathioudakis}, {Doyle},
  {Madjarska}, {Uitenbroek}, \& {Erd{\'e}lyi}}]{2013ApJ...779..125N}
{Nelson}, C.~J., {Shelyag}, S., {Mathioudakis}, M., {et~al.} 2013, \apj, 779,
  125 \csname 2013ApJ...779..125Nlink\endcsname~\csname
  2013ApJ...779..125Nnote\endcsname

\bibitem[{{November} \& {Simon}(1988)}]{1988ApJ...333..427N}
{November}, L.~J. \& {Simon}, G.~W. 1988, \apj, 333, 427 \csname
  1988ApJ...333..427Nlink\endcsname~\csname 1988ApJ...333..427Nnote\endcsname

\bibitem[{{Pariat} {et~al.}(2004){Pariat}, {Aulanier}, {Schmieder},
  {Georgoulis}, {Rust}, \& {Bernasconi}}]{2004ApJ...614.1099P}
{Pariat}, E., {Aulanier}, G., {Schmieder}, B., {et~al.} 2004, \apj, 614, 1099
  \csname 2004ApJ...614.1099Plink\endcsname~\csname
  2004ApJ...614.1099Pnote\endcsname

\bibitem[{{Pariat} {et~al.}(2007){Pariat}, {Schmieder}, {Berlicki}, {Deng},
  {Mein}, {L{\'o}pez Ariste}, \& {Wang}}]{2007A&A...473..279P}
{Pariat}, E., {Schmieder}, B., {Berlicki}, A., {et~al.} 2007, \aap, 473, 279
  \csname 2007A&A...473..279Plink\endcsname~\csname
  2007A&A...473..279Pnote\endcsname

\bibitem[{{Peter} {et~al.}(2014){Peter}, {Tian}, {Curdt}, {Schmit}, {Innes},
  {De Pontieu}, {Lemen}, {Title}, {Boerner}, {Hurlburt}, {Tarbell}, {Wuelser},
  {Mart{\'\i}nez-Sykora}, {Kleint}, {Golub}, {McKillop}, {Reeves}, {Saar},
  {Testa}, {Kankelborg}, {Jaeggli}, {Carlsson}, \&
  {Hansteen}}]{2014Sci...346C.315P}
{Peter}, H., {Tian}, H., {Curdt}, W., {et~al.} 2014, Science, 346, 1255726
  \csname 2014Sci...346C.315Plink\endcsname~\csname
  2014Sci...346C.315Pnote\endcsname

\bibitem[{{Rimmele} {et~al.}(2020){Rimmele}, {Warner}, {Keil}, {Goode},
  {Kn{\"o}lker}, {Kuhn}, {Rosner}, {McMullin}, {Casini}, {Lin}, {W{\"o}ger},
  {von der L{\"u}he}, {Tritschler}, {Davey}, {de Wijn}, {Elmore}, {Fehlmann},
  {Harrington}, {Jaeggli}, {Rast}, {Schad}, {Schmidt}, {Mathioudakis},
  {Mickey}, {Anan}, {Beck}, {Marshall}, {Jeffers}, {Oschmann}, {Beard},
  {Berst}, {Cowan}, {Craig}, {Cross}, {Cummings}, {Donnelly}, {de Vanssay},
  {Eigenbrot}, {Ferayorni}, {Foster}, {Galapon}, {Gedrites}, {Gonzales},
  {Goodrich}, {Gregory}, {Guzman}, {Guzzo}, {Hegwer}, {Hubbard}, {Hubbard},
  {Johansson}, {Johnson}, {Liang}, {Liang}, {McQuillen}, {Mayer}, {Newman},
  {Onodera}, {Phelps}, {Puentes}, {Richards}, {Rimmele}, {Sekulic}, {Shimko},
  {Simison}, {Smith}, {Starman}, {Sueoka}, {Summers}, {Szabo}, {Szabo},
  {Wampler}, {Williams}, \& {White}}]{2020SoPh..295..172R}
{Rimmele}, T.~R., {Warner}, M., {Keil}, S.~L., {et~al.} 2020, \solphys, 295,
  172 \csname 2020SoPh..295..172Rlink\endcsname~\csname
  2020SoPh..295..172Rnote\endcsname

\bibitem[{{Rouppe van der Voort} {et~al.}(2021){Rouppe van der Voort}, {Joshi},
  {Henriques}, \& {Bose}}]{2021A&A...648A..54R}
{Rouppe van der Voort}, L. H.~M., {Joshi}, J., {Henriques}, V. M.~J., \&
  {Bose}, S. 2021, \aap, 648, A54 \csname
  2021A&A...648A..54Rlink\endcsname~\csname 2021A&A...648A..54Rnote\endcsname

\bibitem[{{Rouppe van der Voort} {et~al.}(2016){Rouppe van der Voort},
  {Rutten}, \& {Vissers}}]{2016A&A...592A.100R}
{Rouppe van der Voort}, L. H.~M., {Rutten}, R.~J., \& {Vissers}, G. J.~M. 2016,
  \aap, 592, A100 \csname 2016A&A...592A.100Rlink\endcsname~\csname
  2016A&A...592A.100Rnote\endcsname

\bibitem[{{Roy} \& {Leparskas}(1973)}]{1973SoPh...30..449R}
{Roy}, J.~R. \& {Leparskas}, H. 1973, \solphys, 30, 449 \csname
  1973SoPh...30..449Rlink\endcsname~\csname 1973SoPh...30..449Rnote\endcsname

\bibitem[{{Rutten} {et~al.}(2013){Rutten}, {Vissers}, {Rouppe van der Voort},
  {S{\"u}tterlin}, \& {Vitas}}]{2013JPhCS.440a2007R}
{Rutten}, R.~J., {Vissers}, G. J.~M., {Rouppe van der Voort}, L. H.~M.,
  {S{\"u}tterlin}, P., \& {Vitas}, N. 2013, in Journal of Physics Conference
  Series, Vol. 440, Journal of Physics Conference Series, 012007 \csname
  2013JPhCS.440a2007Rlink\endcsname~\csname 2013JPhCS.440a2007Rnote\endcsname

\bibitem[{{Scharmer} {et~al.}(2003){Scharmer}, {Bjelksj{\"o}}, {Korhonen},
  {Lindberg}, \& {Petterson}}]{2003SPIE.4853..341S}
{Scharmer}, G.~B., {Bjelksj{\"o}}, K., {Korhonen}, T.~K., {Lindberg}, B., \&
  {Petterson}, B. 2003, in \procspie, Vol. 4853, Innovative Telescopes and
  Instrumentation for Solar Astrophysics, ed. S.~L. {Keil} \& S.~V. {Avakyan},
  341--350 \csname 2003SPIE.4853..341Slink\endcsname~\csname
  2003SPIE.4853..341Snote\endcsname

\bibitem[{{Scharmer} {et~al.}(2019){Scharmer}, {L{\"o}fdahl}, {Sliepen}, \& {de
  la Cruz Rodr{\'\i}guez}}]{2019A&A...626A..55S}
{Scharmer}, G.~B., {L{\"o}fdahl}, M.~G., {Sliepen}, G., \& {de la Cruz
  Rodr{\'\i}guez}, J. 2019, \aap, 626, A55 \csname
  2019A&A...626A..55Slink\endcsname~\csname 2019A&A...626A..55Snote\endcsname

\bibitem[{{Scharmer} {et~al.}(2008){Scharmer}, {Narayan}, {Hillberg}, {de la
  Cruz Rodriguez}, {L{\"o}fdahl}, {Kiselman}, {S{\"u}tterlin}, {van Noort}, \&
  {Lagg}}]{2008ApJ...689L..69S}
{Scharmer}, G.~B., {Narayan}, G., {Hillberg}, T., {et~al.} 2008, \apjl, 689,
  L69 \csname 2008ApJ...689L..69Slink\endcsname~\csname
  2008ApJ...689L..69Snote\endcsname

\bibitem[{{Schlichenmaier} {et~al.}(2019){Schlichenmaier}, {Bellot Rubio},
  {Collados}, {Erdelyi}, {Feller}, {Fletcher}, {Jurcak}, {Khomenko},
  {Leenaarts}, {Matthews}, {Belluzzi}, {Carlsson}, {Dalmasse}, {Danilovic},
  {G{\"o}m{\"o}ry}, {Kuckein}, {Manso Sainz}, {Martinez Gonzalez},
  {Mathioudakis}, {Ortiz}, {Riethm{\"u}ller}, {Rouppe van der Voort}, {Simoes},
  {Trujillo Bueno}, {Utz}, \& {Zuccarello}}]{2019arXiv191208650S}
{Schlichenmaier}, R., {Bellot Rubio}, L.~R., {Collados}, M., {et~al.} 2019,
  arXiv e-prints, arXiv:1912.08650 \csname
  2019arXiv191208650Slink\endcsname~\csname 2019arXiv191208650Snote\endcsname

\bibitem[{{Severny}(1964)}]{1964ARA&A...2..363S}
{Severny}, A.~B. 1964, \araa, 2, 363 \csname
  1964ARA&A...2..363Slink\endcsname~\csname 1964ARA&A...2..363Snote\endcsname

\bibitem[{{Shetye} {et~al.}(2018){Shetye}, {Shelyag}, {Reid}, {Scullion},
  {Doyle}, \& {Arber}}]{2018MNRAS.479.3274S}
{Shetye}, J., {Shelyag}, S., {Reid}, A.~L., {et~al.} 2018, \mnras, 479, 3274
  \csname 2018MNRAS.479.3274Slink\endcsname~\csname
  2018MNRAS.479.3274Snote\endcsname

\bibitem[{{Spruit} {et~al.}(1990){Spruit}, {Nordlund}, \&
  {Title}}]{1990ARA&A..28..263S}
{Spruit}, H.~C., {Nordlund}, A., \& {Title}, A.~M. 1990, \araa, 28, 263 \csname
  1990ARA&A..28..263Slink\endcsname~\csname 1990ARA&A..28..263Snote\endcsname

\bibitem[{{van Noort} {et~al.}(2005){van Noort}, {Rouppe van der Voort}, \&
  {L{\"o}fdahl}}]{2005SoPh..228..191V}
{van Noort}, M., {Rouppe van der Voort}, L., \& {L{\"o}fdahl}, M.~G. 2005,
  \solphys, 228, 191 \csname 2005SoPh..228..191Vlink\endcsname~\csname
  2005SoPh..228..191Vnote\endcsname

\bibitem[{{Vissers} {et~al.}(2013){Vissers}, {Rouppe van der Voort}, \&
  {Rutten}}]{2013ApJ...774...32V}
{Vissers}, G.~J.~M., {Rouppe van der Voort}, L.~H.~M., \& {Rutten}, R.~J. 2013,
  \apj, 774, 32 \csname 2013ApJ...774...32Vlink\endcsname~\csname
  2013ApJ...774...32Vnote\endcsname

\bibitem[{{Vissers} {et~al.}(2019){Vissers}, {Rouppe van der Voort}, \&
  {Rutten}}]{2019A&A...626A...4V}
{Vissers}, G. J.~M., {Rouppe van der Voort}, L. H.~M., \& {Rutten}, R.~J. 2019,
  \aap, 626, A4 \csname 2019A&A...626A...4Vlink\endcsname~\csname
  2019A&A...626A...4Vnote\endcsname

\bibitem[{{Vissers} {et~al.}(2015){Vissers}, {Rouppe van der Voort}, {Rutten},
  {Carlsson}, \& {De Pontieu}}]{2015ApJ...812...11V}
{Vissers}, G.~J.~M., {Rouppe van der Voort}, L.~H.~M., {Rutten}, R.~J.,
  {Carlsson}, M., \& {De Pontieu}, B. 2015, \apj, 812, 11 \csname
  2015ApJ...812...11Vlink\endcsname~\csname 2015ApJ...812...11Vnote\endcsname

\bibitem[{{Wang}(1989)}]{1989SoPh..123...21W}
{Wang}, H. 1989, \solphys, 123, 21 \csname
  1989SoPh..123...21Wlink\endcsname~\csname 1989SoPh..123...21Wnote\endcsname

\bibitem[{{Watanabe} {et~al.}(2008){Watanabe}, {Kitai}, {Okamoto}, {Nishida},
  {Kiyohara}, {Ueno}, {Hagino}, {Ishii}, \& {Shibata}}]{2008ApJ...684..736W}
{Watanabe}, H., {Kitai}, R., {Okamoto}, K., {et~al.} 2008, \apj, 684, 736
  \csname 2008ApJ...684..736Wlink\endcsname~\csname
  2008ApJ...684..736Wnote\endcsname

\bibitem[{{Watanabe} {et~al.}(2011){Watanabe}, {Vissers}, {Kitai}, {Rouppe van
  der Voort}, \& {Rutten}}]{2011ApJ...736...71W}
{Watanabe}, H., {Vissers}, G., {Kitai}, R., {Rouppe van der Voort}, L., \&
  {Rutten}, R.~J. 2011, \apj, 736, 71 \csname
  2011ApJ...736...71Wlink\endcsname~\csname 2011ApJ...736...71Wnote\endcsname

\bibitem[{{Yelles Chaouche} {et~al.}(2011){Yelles Chaouche}, {Moreno-Insertis},
  {Mart{\'\i}nez Pillet}, {Wiegelmann}, {Bonet}, {Kn{\"o}lker}, {Bellot Rubio},
  {del Toro Iniesta}, {Barthol}, {Gandorfer}, {Schmidt}, \&
  {Solanki}}]{2011ApJ...727L..30Y}
{Yelles Chaouche}, L., {Moreno-Insertis}, F., {Mart{\'\i}nez Pillet}, V.,
  {et~al.} 2011, \apjl, 727, L30 \csname
  2011ApJ...727L..30Ylink\endcsname~\csname 2011ApJ...727L..30Ynote\endcsname

\bibitem[{{Young} {et~al.}(2018){Young}, {Tian}, {Peter}, {Rutten}, {Nelson},
  {Huang}, {Schmieder}, {Vissers}, {Toriumi}, {Rouppe van der Voort},
  {Madjarska}, {Danilovic}, {Berlicki}, {Chitta}, {Cheung}, {Madsen},
  {Reardon}, {Katsukawa}, \& {Heinzel}}]{2018SSRv..214..120Y}
{Young}, P.~R., {Tian}, H., {Peter}, H., {et~al.} 2018, \ssr, 214, 120 \csname
  2018SSRv..214..120Ylink\endcsname~\csname 2018SSRv..214..120Ynote\endcsname

\end{thebibliography}

\begin{appendix}

\section{$k$-means clustering}
\label{app:kmeans}

\begin{figure}
    \includegraphics[width=0.48\textwidth]{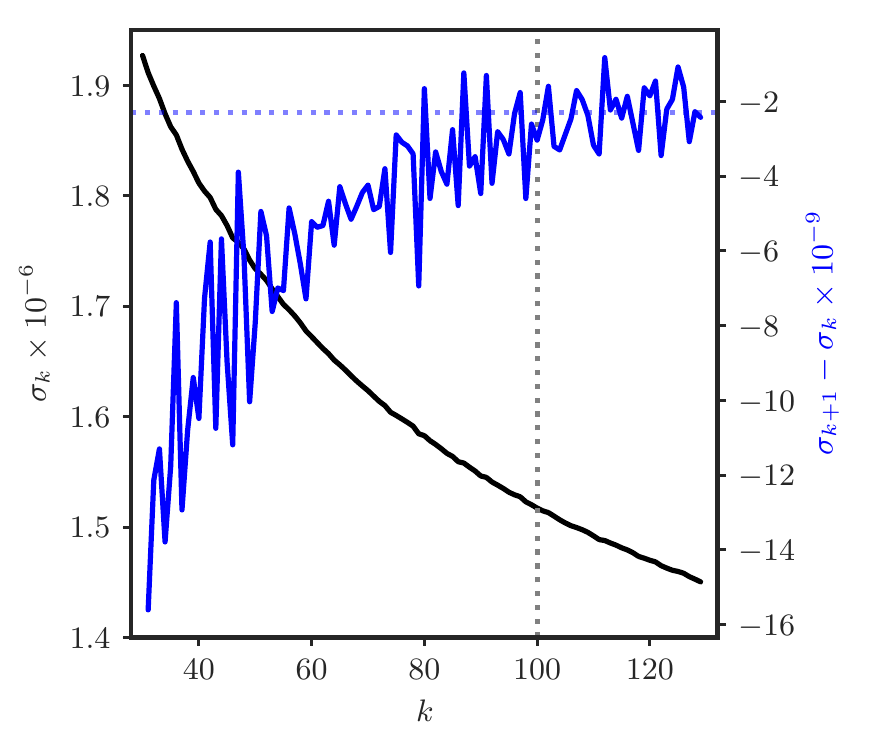}
    \caption{\label{fig:no_cluster}%
    Finding the number of clusters, $k$, for optimally clustering the \Hbeta{} spectra. Variations in the inertia ($\sigma_{k}$) with respect to $k$ is shown in black. The presented $\sigma_{k}$ is normalized with the total number of data points used in the training of the $k$-means model. The running difference,  $\sigma_{k+1}-\sigma_{k}$ is plotted in blue. The vertical dotted black line indicates the used number of clusters, $k=100$, for the final clustering.}
\end{figure}

The unsupervised $k$-means algorithm requires a predetermined cluster number, $k$, to partition the data. 
Selecting the optimal number of clusters is a crucial step for efficient application of the $k$-means clustering. 
We used the elbow method to determine the number of clusters, where we analyze the change in the total inertia ($\sigma_{k}$, within-cluster sum of squares of euclidean distances) with respect to varying $k$ from 30 to 130. 
The variations in $\sigma_{k}$ divided by the total number of pixels used in the training of the $k$-means algorithm is presented in Fig.~\ref{fig:no_cluster}.    
The total inertia decreases with increase in $k$. 
In principle, $\sigma_{k}$ should reach its minimum value when $k$ is equal to total number of pixels. However, the purpose of the $k$-means method is to reduce data points into meaningful clusters to facilities an efficient data analysis.
The elbow method finds a certain $k$ value after which $\sigma_{k}$ decreases linearly.
It is evident that $\sigma_{k}$ decrease almost linear for $k$ higher than 100, which can been seen in variations of  $\sigma_{k+1}-\sigma_{k}$ plotted in Fig.~\ref{fig:no_cluster}.
Therefore, we choose $k=100$ for the $k$-means clustering of the \Hbeta{} spectra. 

\begin{figure*}
    \includegraphics[width=\textwidth]{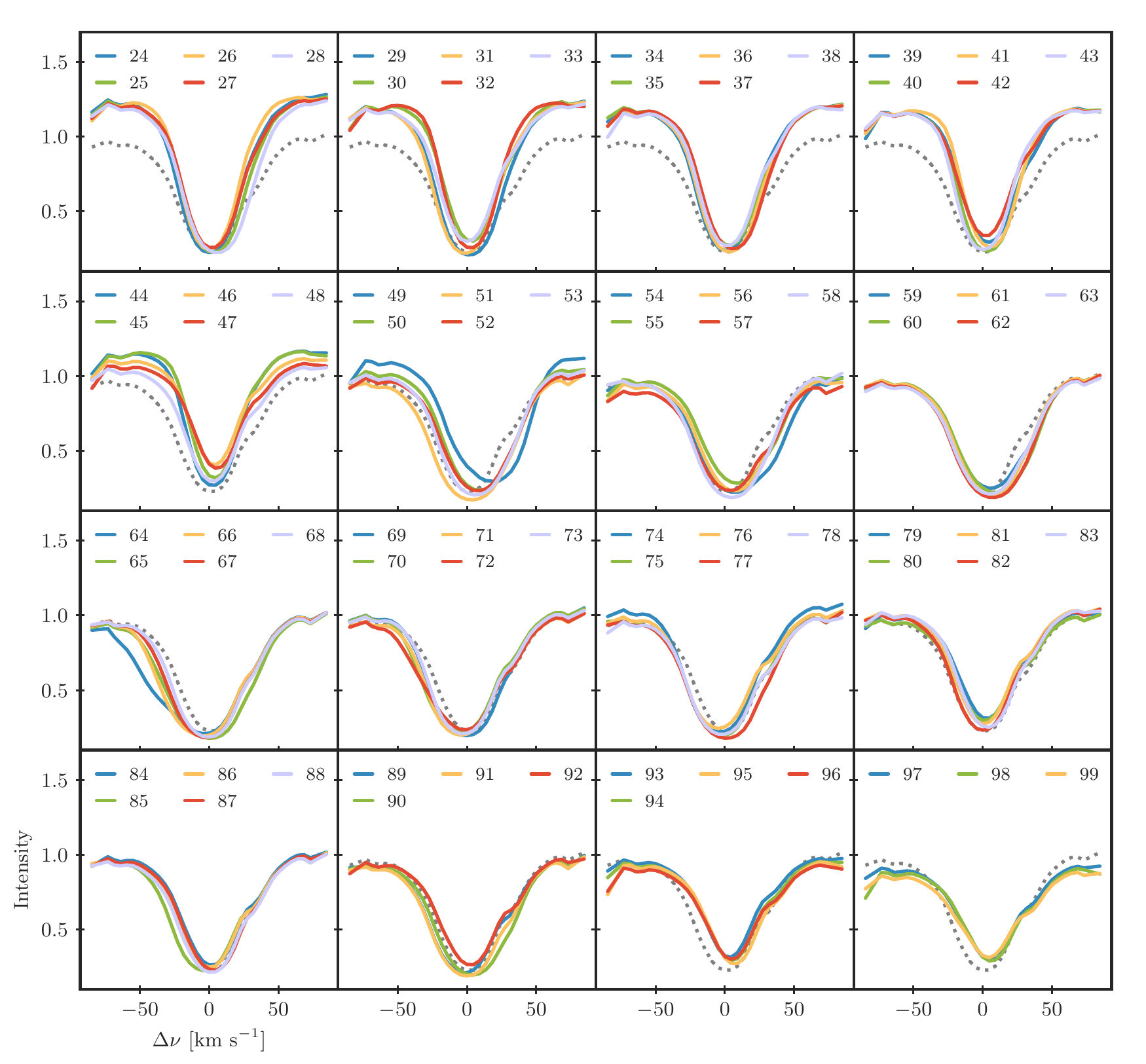}
    \caption{\label{fig:rest_RPs}%
    Representative profiles obtained through $k$-means clustering of the \Hbeta{} spectra. The remaining 76 RPs that were not considered as QSEB profiles are shown in 16 panels (RPs 24--99). The QSEB RPs (0--23) are shown in Fig.~\ref{fig:RPs}. 
    The grey dashed profile in each panel represents the average quiet Sun profile. 
    }
\end{figure*}

Representative profiles 0--23 selected for the detection of QSEBs are shown in Fig.~\ref{fig:RPs} and the remaining RPs (RP 24--99) are presented in Fig.~\ref{fig:rest_RPs}.
 
\end{appendix}

\end{document}